\documentclass[a4paper, 11pt, oneside]{article}
\pdfoutput=1
\usepackage{float}
\usepackage{jheppub}
\usepackage{epstopdf}

\usepackage{bbm}   
\usepackage{amsmath}
\usepackage{graphicx}

\newcommand{\nc}{\newcommand}

\newlength{\absize}
\nc{\non}{\nonumber}
\nc{\hc}{\hbox {H.c.}} 
\nc{\noi}{\noindent}
\nc{\barx}{\bar{x}}
\nc{\pbarn}{\;\hbox {pb}}
\nc{\fbarn}{\;\hbox {fb}}
\newcommand{\bi}{\begin{itemize}}
\newcommand{\ei}{\end{itemize}}
\newcommand{\lam}{\lambda}
\def\thetaW{{\theta}_\text{W}}
\nc{\lsp}{\;\;\;\;\;}
\nc{\Lsp}{\;\;\;\;\;\;\;\;\;\;}  
\nc{\LLsp}{\lspace \lspace}
\nc{\lra}{\longrightarrow}
\nc{\beq}{\begin{equation}}  \nc{\eeq}{\end{equation}}
\nc{\bea}{\begin{eqnarray}}  \nc{\eea}{\end{eqnarray}}
\nc{\baa}{\begin{array}}     \nc{\eaa}{\end{array}}
\nc{\bit}{\begin{itemize}}   \nc{\eit}{\end{itemize}}
\nc{\ben}{\begin{enumerate}} \nc{\een}{\end{enumerate}}
\nc{\bce}{\begin{center}}    \nc{\ece}{\end{center}}
\nc{\bpm}{\begin{pmatrix}}   \nc{\epm}{\end{pmatrix}}
\nc{\bvt}{\begin{verbatim}}  \nc{\evt}{\end{verbatim}}
\def\lsim{\mathrel{\raise.3ex\hbox{$<$\kern-.75em\lower1ex\hbox{$\sim$}}}}
\def\gsim{\mathrel{\raise.3ex\hbox{$>$\kern-.75em\lower1ex\hbox{$\sim$}}}}

\def\mcal{{\cal M}}

\def\pcal{{\cal P}}

\def\gev{\;\hbox{GeV}}

\def\zBB{{\mathbbm Z}}

\nc{\tanb}{\tan\beta}
\nc{\mch}{M_{H^\pm}}

\def\cb{c_\beta}
\def\sb{s_\beta}
\def\tb{t_\beta}
\def\mch{M_{H^\pm}}

\nc{\for}{\lsp {\rm for} \lsp}
\nc{\andd}{\lsp {\rm and} \lsp}

\renewcommand{\Re}{\mbox{Re\thinspace}}
\renewcommand{\Im}{\mbox{Im\thinspace}}
\newcommand{\half}{{\textstyle\frac{1}{2}}}
\allowdisplaybreaks %!!!!

\def\i11{{\mathbbm 1}}%\def\iBB{ \hbox{{\mysmallii I}}\!\hbox{{\mysmallii I}} }

\title{Measuring CP violation in Two-Higgs-Doublet models \\ in light of the LHC Higgs data}
\author[a]{B. Grzadkowski,}
\affiliation[a]{Faculty of Physics, University of Warsaw, Ho\.za 69, 00-681 Warsaw, Poland}
\author[b]{O. M. Ogreid,}
\affiliation[b]{Bergen University College, Postboks 7030, N-5020 Bergen, Norway}
\author[c]{P. Osland}
\affiliation[c]{Department of Physics,
University of Bergen, Postboks 7803, N-5020 Bergen, Norway}
\emailAdd{bohdan.grzadkowski@fuw.edu.pl}
\emailAdd{omo@hib.no}
\emailAdd{Per.Osland@ift.uib.no}
\date{\today}

\abstract{In Two-Higgs-Doublet models, the conditions for CP violation can be expressed in terms of invariants under U(2) rotations among the two SU(2) Higgs doublet fields. In order to design a strategy for measuring the invariants we express them in terms of observables, i.e., masses and couplings of scalar bosons. We find amplitudes directly sensitive to the invariants.
Observation of the Standard-Model-like Higgs boson at the LHC severely constrains the models.
In particular, in the model with $\zBB_2$ symmetry imposed on
dimension-4 terms (in order to eliminate tree-level flavour-changing
neutral currents), CP violation is strongly suppressed.
On the other hand, the most general Two-Higgs-Doublet model (without $\zBB_2$ symmetry) is compatible with the LHC data, and would still allow for CP violation to be present in the model.
Consequently, also flavour-changing neutral currents would in general be expected.
We briefly sketch a strategy for measuring the remaining CP violation.}


\begin{document}

\maketitle

\flushbottom

%%%%%%%%%%%%%%%%%%%%%%%%%%%%%%%%%%%%%%%%%%%%%%%%%%%%%%%%%%%%%%%%%%%%%%%%%%%%%
\section{Introduction}
\label{Sec:Introduction}
%%%%%%%%%%%%%%%%%%%%%%%%%%%%%%%%%%%%%%%%%%%%%%%%%%%%%%%%%%%%%%%%%%%%%%%%%%%%%

As is well known, the Two-Higgs-Doublet model (2HDM) allows for extra sources of CP violation
that originate in the scalar potential. 
This possibility opens interesting perspectives for cosmology \cite{Riotto:1999yt}.
Nevertheless, this option receives little attention in much of the literature \cite{Branco:2011iw}. 
The 2HDM can be formulated in any basis chosen for the Higgs doublets, e.g.\
any U(2) rotation acting upon the doublets $(\Phi_1,\Phi_2)$ would lead to an equivalent
basis. Thus, physical implications, such as cross sections, decay widths etc., 
can only depend on quantities that are basis independent.
In this paper we discuss CP violation originating from the scalar sector of the 2HDM. In order
to present results in a basis independent way, we are going to study weak basis invariants
sensitive to CP violation.

Necessary conditions for having CP violation in this model were first formulated in terms of invariants 20 years ago by 
Lavoura, Silva and Botella \cite{Lavoura:1994fv,Botella:1994cs}. More recently, this
issue was addressed by Branco, Rebelo and Silva-Marcos \cite{Branco:2005em}, by Gunion and Haber \cite{Gunion:2005ja}
and by Haber and O'Neil~\cite{Haber:2006ue}.
Independent approaches have been presented both in terms of algebraic invariants \cite{Davidson:2005cw} 
and geometric quantities \cite{Ivanov:2005hg,Nishi:2006tg,Ivanov:2006yq,Maniatis:2007vn}.
Detailed discussions of CP-violating invariants are also contained in \cite{Branco:1999fs}.
These invariants are analogous to the Jarlskog invariant $J$ \cite{Jarlskog:1985ht} 
describing CP violation induced by the Yukawa couplings in the Standard Model (SM).

There exist several versions of the 2HDM which differ by the Yukawa interactions, e.g. type I or type II 2HDM.
Our intention in this paper is to present results which are type-independent, thus insensitive
to the Yukawa structure, and hence applicable in any 2HDM. Therefore we are going to restrict 
ourselves to the bosonic sector of the model. The Yukawa sector will in general supply additional sources of CP violation.

The paper is organized as follows. In section~\ref{Sec:def-model} we review the model, and establish our notation. 
In section~\ref{Sec:CPviolation} we discuss CP violation, and present the criteria for CP violation in 
terms of physical couplings and masses.
In section~\ref{Sec:INV} we relate these invariants to physical amplitudes involving scalars and vector bosons. 
Then, in section~\ref{Sec:H1SM limit} we discuss the limit in which the 125~GeV Higgs particle observed 
at the LHC \cite{Aad:2012tfa,Chatrchyan:2012ufa} is the lightest neutral Higgs boson of the model, 
and couples to vector bosons like the SM Higgs boson. We show that the most general 2HDM still allows for CP violation involving the heavier companions, and also that tree-level flavour violation in couplings of neutral scalars could be present. In section~\ref{Sec:Illustrations} we present some numerical illustrations, in section~\ref{Sec:Strategy} we outline a strategy for systematically excluding or
discovering CP violation in the model, and in section~\ref{Sec:Summary} 
we summarize our main points. Technical details are relegated to two appendices.

%%%%%%%%%%%%%%%%%%%%%%%%%%%%%%%%%%%%%%%%%%%%%%%%%%%%%%%%%%%%%%%%%%%%%%%%%%%%%
\section{The model}
\label{Sec:def-model}
\setcounter{equation}{0}
%%%%%%%%%%%%%%%%%%%%%%%%%%%%%%%%%%%%%%%%%%%%%%%%%%%%%%%%%%%%%%%%%%%%%%%%%%%%%

The scalar potential of the 2HDM shall be parametrized in the standard fashion:
\begin{align}
\label{Eq:pot}
V(\Phi_1,\Phi_2) &= -\frac12\left\{m_{11}^2\Phi_1^\dagger\Phi_1
+ m_{22}^2\Phi_2^\dagger\Phi_2 + \left[m_{12}^2 \Phi_1^\dagger \Phi_2
+ \hc\right]\right\} \nonumber \\
& + \frac{\lambda_1}{2}(\Phi_1^\dagger\Phi_1)^2
+ \frac{\lambda_2}{2}(\Phi_2^\dagger\Phi_2)^2
+ \lambda_3(\Phi_1^\dagger\Phi_1)(\Phi_2^\dagger\Phi_2) 
+ \lambda_4(\Phi_1^\dagger\Phi_2)(\Phi_2^\dagger\Phi_1)\nonumber \\
&+ \frac12\left[\lambda_5(\Phi_1^\dagger\Phi_2)^2 + \hc\right]
+\left\{\left[\lambda_6(\Phi_1^\dagger\Phi_1)+\lambda_7
(\Phi_2^\dagger\Phi_2)\right](\Phi_1^\dagger\Phi_2)
+{\rm \hc}\right\} \\
&\equiv Y_{a\bar{b}}\Phi_{\bar{a}}^\dagger\Phi_b+\frac{1}{2}Z_{a\bar{b}c\bar{d}}(\Phi_{\bar{a}}^\dagger\Phi_b)(\Phi_{\bar{c}}^\dagger\Phi_d).
\label{Eq:pot-other}
\end{align}
In the second form, Eq.~(\ref{Eq:pot-other}), a summation over barred with un-barred indices is implied, e.g., $a=\bar a=1,2$. Thus,
\bea
Y_{11}=-\frac{m_{11}^2}{2},\quad Y_{12}=-\frac{m_{12}^2}{2},\quad  Y_{21}=-\frac{(m_{12}^2)^*}{2},\quad Y_{22}=-\frac{m_{22}^2}{2}
\eea
and
\bea
&&Z_{1111}=\lambda_1,\quad Z_{2222}=\lambda_2,\quad Z_{1122}=Z_{2211}=\lambda_3,\nonumber\\
&&Z_{1221}=Z_{2112}=\lambda_4,\quad Z_{1212}=\lambda_5,\quad Z_{2121}=(\lambda_5)^*,\nonumber\\
&&Z_{1112}=Z_{1211}=\lambda_6,\quad Z_{1121}=Z_{2111}=(\lambda_6)^*,\nonumber\\
&&Z_{1222}=Z_{2212}=\lambda_7,\quad Z_{2122}=Z_{2221}=(\lambda_7)^*.
\eea
All other $Z_{a\bar{b}c\bar{d}}$ vanish.

Usually a $\zBB_2$ symmetry is imposed on the dimension-4 terms in order to eliminate potentially large flavour-changing neutral currents in the Yukawa couplings. We will in the present work not restrict ourselves by imposing this symmetry, and therefore we are going to consider the most
general scalar potential, keeping also terms that are not allowed by $\zBB_2$ symmetry.

In an arbitrary basis, the vacuum may be complex, and the Higgs doublets can be parameterized as 
\begin{equation}
\Phi_j=e^{i\xi_j}\left(
\begin{array}{c}\varphi_j^+\\ (v_j+\eta_j+i\chi_j)/\sqrt{2}
\end{array}\right), \quad
j=1,2.\label{vevs}
\end{equation}
Here $v_j$ are real numbers, so that $v_1^2+v_2^2=v^2$. The fields $\eta_j$ and $\chi_j$ are real.
 The phase difference between the two vevs is given by
\beq
\xi\equiv\xi_2-\xi_1.
\eeq
The vevs may also be written as
\begin{equation}
\left<\Phi_j\right>=\frac{1}{\sqrt{2}}\left(
\begin{array}{c}0\\ v\hat{v}_j
\end{array}\right).
\end{equation}
where
\bea
\hat{v}_1=\frac{v_1}{v}e^{i\xi_1},\quad \hat{v}_2=\frac{v_2}{v}e^{i\xi_2},
\eea
which will be useful later.
Next, let's define orthogonal states
\beq
\left(
\begin{array}{c}G_0\\ \eta_3
\end{array}\right)
=
\left(
\begin{array}{cc}v_1/v & v_2/v\\ -v_2/v & v_1/v
\end{array}\right)
\left(
\begin{array}{c}\chi_1\\ \chi_2
\end{array}\right)
\eeq
and
\beq
\left(
\begin{array}{c}G^\pm\\ H^\pm
\end{array}\right)
=
\left(
\begin{array}{cc}v_1/v & v_2/v\\ -v_2/v & v_1/v
\end{array}\right)
\left(
\begin{array}{c}\varphi_1^\pm\\ \varphi_2^\pm
\end{array}\right).
\eeq
Then $G_0$ and $G^\pm$ become the massless Goldstone fields, and $H^\pm$ are the charged scalars.

The model also contains three neutral scalars, which are linear compositions of the $\eta_i$,
\begin{equation} \label{Eq:R-def}
\begin{pmatrix}
H_1 \\ H_2 \\ H_3
\end{pmatrix}
=R
\begin{pmatrix}
\eta_1 \\ \eta_2 \\ \eta_3
\end{pmatrix},
\end{equation}
with the $3\times3$ orthogonal rotation matrix $R$ satisfying
\begin{equation}
\label{Eq:cal-M}
R{\cal M}^2R^{\rm T}={\cal M}^2_{\rm diag}={\rm diag}(M_1^2,M_2^2,M_3^2),
\end{equation}
and with $M_1\leq M_2\leq M_3$. A convenient parametrization of the rotation matrix $R$ is \cite{Accomando:2006ga,El_Kaffas:2006nt}
\begin{equation} 
R=
\begin{pmatrix}
R_{11}    &  R_{12}   & R_{13}   \\
R_{21}    &  R_{22}   & R_{23}   \\
R_{31}    &  R_{32}   & R_{33}  
\end{pmatrix}
=
\begin{pmatrix}
c_1\,c_2 & s_1\,c_2 & s_2 \\
- (c_1\,s_2\,s_3 + s_1\,c_3) 
& c_1\,c_3 - s_1\,s_2\,s_3 & c_2\,s_3 \\
- c_1\,s_2\,c_3 + s_1\,s_3 
& - (c_1\,s_3 + s_1\,s_2\,c_3) & c_2\,c_3
\end{pmatrix}.
\end{equation}
Since $R$ is orthogonal, only three of the elements $R_{ij}$ are independent. The rest can be expressed by these through the use of orthogonality relations.
From the potential one can now derive expressions for the masses of the scalars as well as Feynman rules for  scalar interactions. For a general basis that we consider here,
these expressions are quite involved and lengthy so we have chosen to collect them in Appendix \ref{Sec:MoreProperties}.

%%%%%%%%%%%%%%%%%%%%%%%%%%%%%%%%%%%%%%%%%%%%%%%%%%%%%%%%%%%%%%%%%%%%%%%%%%%%%
\section{CP violation}
\label{Sec:CPviolation}
\setcounter{equation}{0}
%%%%%%%%%%%%%%%%%%%%%%%%%%%%%%%%%%%%%%%%%%%%%%%%%%%%%%%%%%%%%%%%%%%%%%%%%%%%%

The addition of the second doublet triggers qualitatively new phenomena originating
from interactions of scalar particles. The crucial one is the attractive possibility
of CP violation in the scalar potential \cite{Lee:1973iz}. This extra source of CP violation
might be very essential for explaining the baryon asymmetry. In this section we are going
to discuss parametrization of CP violation in terms of weak-basis invariants.

%%%%%%%%%%%%%%%%%%%%%%%%%%%%%%%%%%%%%%%%%%%%%%%%%%%%%%%%%%%%%%%%%%%%%%%%%%%%%
\subsection{Conditions for CP violation}
%%%%%%%%%%%%%%%%%%%%%%%%%%%%%%%%%%%%%%%%%%%%%%%%%%%%%%%%%%%%%%%%%%%%%%%%%%%%%

As pointed out by Gunion and Haber, the conditions for having CP violation in the model can be expressed in terms of three U(2) invariants constructed from coefficients of the quadratic ($Y$) and dimension-4 ($Z$) terms of the potential, together with the vacuum expectation values. 

It was found \cite{Gunion:2005ja} that in order to break CP at least one of following three invariants had to be non-zero:
\begin{subequations}
\begin{align} \label{eq:im_J1}
\Im J_1&=-\frac{2}{v^2}\Im\bigl[\hat{v}_{\bar{a}}^* Y_{a\bar{b}} Z_{b\bar{d}}^{(1)}\hat{v}_d\bigr], \\
\label{eq:im_J2}
\Im J_2&=\frac{4}{v^4}\Im\bigl[\hat{v}_{\bar{b}}^* \hat{v}_{\bar{c}}^* Y_{b\bar{e}} Y_{c\bar{f}} Z_{e\bar{a}f\bar{d}}\hat{v}_a\hat{v}_d\bigr], \\
\Im J_3&=\Im\bigl[\hat{v}_{\bar{b}}^* \hat{v}_{\bar{c}}^* Z_{b\bar{e}}^{(1)} Z_{c\bar{f}}^{(1)}Z_{e\bar{a}f\bar{d}}\hat{v}_a\hat{v}_d\bigr],
 \label{eq:im_J3}
\end{align}
\end{subequations}
where $Z_{a\bar{d}}^{(1)}\equiv\delta_{b\bar{c}}Z_{a\bar{b}c\bar{d}}$. 
Having invariants expressed by the parameters of the potential and the vevs, 
one is faced with the challenge of measuring all these parameters in order to determine the CP properties of the model \cite{Grzadkowski:2013rza}.
For this purpose it would be much more convenient to formulate the conditions for CP violation in terms of physically measurable quantities like masses and couplings.
%%%%%%%%%%%%%%%%%%%%%%%%%%%%%%%%%%%%%%%%%%%%%%%%%%%%%%%%%%%%%%%%%%%%%%%%%%%%%
\subsection{Expressing $\Im J_i$ in terms of masses and couplings}
%%%%%%%%%%%%%%%%%%%%%%%%%%%%%%%%%%%%%%%%%%%%%%%%%%%%%%%%%%%%%%%%%%%%%%%%%%%%%
Our aim is to express $\Im J_i,$  $i=1,2,3$ in terms of physical quantities. We shall start by first writing out these expressions explicitly in terms of the parameters of the potential and the vevs. Then we will re-express the original parameters of the potential in terms of another set of parameters\footnote{The potential (\ref{Eq:pot-other}) contains 14 real parameters. 
However, it is worth realizing that by an appropriate choice of basis, one can reduce the number of free parameters to 11. 
Nevertheless, in order to preserve and control the invariance with respect to basis transformations, hereafter we keep the set of 14 parameters unless explicitly stated otherwise.}
:
\bea 
\pcal_{67}\equiv\{M_{H^\pm}^2,\mu^2,M_1^2,M_2^2,M_3^2,{\rm Im}\lambda_5,{\rm Re}\lambda_6,{\rm Re}\lambda_7, v_1, v_2, \xi,\alpha_1,\alpha_2,\alpha_3\}.\label{input}
\eea
See  Appendix~\ref{Sec:MoreProperties} for details.

The resulting expressions are large, but can be handled efficiently by computer algebra. We used {\tt Mathematica} \cite{mathematica} for this purpose.
Using (\ref{stationary1})--(\ref{stationary3}) along with (\ref{l1rephrased})--(\ref{iml7rephrased}), we 
are able to express  $\Im J_i$ in terms of the parameters $\pcal_{67}$ listed in (\ref{input}). 
It is also worth noticing that each of the  $\Im J_i$ is a homogeneous polynomial in a subset of $\pcal_{67}$ defined as 
\begin{equation} \label{eq:pcal0}
\pcal_0=\{M_{H^\pm}^2,\mu^2,M_1^2,M_2^2,M_3^2,{\rm Im}\lambda_5,{\rm Re}\lambda_6,{\rm Re}\lambda_7\},
\end{equation}
for details see (\ref{l1rephrased})--(\ref{iml7rephrased}). 
In particular,  $\Im J_1$ is of order 2, whereas $\Im J_2$ and $\Im J_3$ are of order 3. This means that by expanding these expressions in the parameters of $\pcal_0$, we get 36 terms in the expansion of $\Im J_1$ and 120 terms in the expansions of $\Im J_2$ and $\Im J_3$.

We denote the couplings $(H_i VV)$, $(H_iH^+H^-)$ and $(H^+H^-H^+H^-)$, by $e_i$, $q_i$ and $q$, respectively, details 
are contained in Appendix \ref{Sec:Couplings}. 

Let us start by investigating $\Im J_2$, since this turns out to be the simplest of the three invariants. By writing out all the 120 terms of this invariant, using the orthogonality of the rotation matrix, we find that the terms containing $M_{H^\pm}^2,\mu^2,{\rm Im}\lambda_5,{\rm Re}\lambda_6,{\rm Re}\lambda_7$ all vanish. The resulting expression becomes 
\bea
\Im J_2&=&\frac{2e_1 e_2 e_3}{v^9}(M_1^2-M_2^2)(M_2^2-M_3^2)(M_3^2-M_1^2)\nonumber\\
&=&\frac{2}{v^9}\sum_{i,j,k}\epsilon_{ijk}e_ie_je_kM_i^4M_k^2=\frac{2e_1 e_2 e_3}{v^9}\sum_{i,j,k}\epsilon_{ijk}M_i^4M_k^2.
\eea
We note that this expression is completely antisymmetric under the interchange of two of the indices $i,j,k$, labeling the three neutral Higgs fields. 
The above formula found in the general basis confirms the result obtained in the ``Higgs basis" 
($v_1=v$ and $v_2=0$) in \cite{Lavoura:1994fv}.

Next, we turn to $\Im J_1$. By writing out all the 36 terms of this invariant, using the orthogonality of the rotation matrix, we find that terms not containing neutral Higgs masses vanish. Also, the terms containing $M_i^4$ vanish in this expansion. Inspired by the results for $\Im J_2$, we conjectured that also the expression for $\Im J_1$ should be completely antisymmetric under the exchange of two of the indices $i,j,k$. A careful study of the coefficients of $\Im J_1$ in the expansion of parameters of $\pcal_0$ suggests we look for an expression proportional to $\sum_{i,j,k}\epsilon_{ijk}e_i^ae_j^be_k^cM_i^2q_j$ which is of order 2 in the parameters of $\pcal_0$. Under this conjecture, we tried out different small values of $a,b,c$, seeing if we could reproduce the expression for $\Im J_1$. After a small game of trial and error we hit the jackpot by putting $a=c=1$ and $b=0$, establishing the relation
\bea
\Im J_1&=&\frac{1}{v^5}\sum_{i,j,k}\epsilon_{ijk}M_i^2e_ie_kq_j\nonumber\\
&=&\frac{1}{v^5}[M_1^2e_1(e_3q_2-e_2q_3)+M_2^2e_2(e_1q_3-e_3q_1)+M_3^2e_3(e_2q_1-e_1q_2)].
\eea

$\Im J_3$ turns out to be more complex. In fact it contains some terms with $\Im J_1$ and $\Im J_2$ plus  ``independent" terms.
We have established the following identity
\begin{equation}
\Im J_3= K\,\Im J_1 +\Im J_2
+\frac{2}{v^7}\sum_{i,j,k}\epsilon_{ijk}(v^2 q_i +2e_i M_i^2)M_i^2  e_j q_k
\end{equation}
with
\begin{align}
K&=\frac{2}{v^4}\left[ (e_1^2M_1^2+e_2^2M_2^2+e_3^2M_3^2)
+v^2(e_1q_1  + e_2q_2 +  e_3q_3) +2v^2M_{H^\pm}^2\right] \nonumber \\
&=\frac{2}{v^4}\left[(e_1^2M_1^2+e_2^2M_2^2+e_3^2M_3^2)+2v^4\sigma-4v^4 q
+2v^2M_{H^\pm}^2\right],
\end{align}
where $\sigma$ and $q$ are defined in equations (\ref{eq:sigma}) and (\ref{Eq:coupling-q}).

By putting all the $\Im J_i=0$ and solving the resulting three equations, we arrive at six distinct cases under which we have CP conservation:\\
\\
Case 1: $M_1=M_2=M_3$. Full mass degeneracy.\\
Case 2: $M_1=M_2$ and $e_1q_2 = e_2q_1$.\\
Case 3: $M_2=M_3$ and $e_2q_3 = e_3q_2$.\\
Case 4: $e_1=0$ and $q_1=0$.\\
Case 5: $e_2=0$ and $q_2=0$.\\
Case 6: $e_3=0$ and $q_3=0$.\\
\\
The obvious solution $e_1=e_2=e_3=0$ is not included since $e_1^2+e_2^2+e_3^2=v^2\neq0$, and this solution would be unphysical.
If one (or more) of the 6 above cases occur, it means that the 2HDM is CP conserving. If none of the above cases occurs, 
it means that the 2HDM violates CP. It is worth noticing that the
nature of CP violation is not revealed at this point,
i.e., CP could be broken explicitly or spontaneously~\cite{Grzadkowski:2013rza}.
%%%%%%%%%%%%%%%%%%%%%%%%%%%%%%%%%%%%%%%%%%%%%%%%%%%%%%%%%%%%%%%%%%%%%%%%%%%%%
\subsection{Re-expressing the conditions for CP violation}
%%%%%%%%%%%%%%%%%%%%%%%%%%%%%%%%%%%%%%%%%%%%%%%%%%%%%%%%%%%%%%%%%%%%%%%%%%%%%
While $\Im J_1$ and $\Im J_2$ are somewhat ``atomic" in form when written as an antisymmetric sum, $\Im J_3$ is not. 
Let us therefore focus on the ``independent" terms in the expression for $\Im J_3$, and split them like
\bea
\frac{2}{v^7}\sum_{i,j,k}\epsilon_{ijk}(v^2 q_i +2e_i M_i^2)M_i^2  e_j q_k=4\Im J_{10}+2\Im J_{30}.
\eea
where we have put
\bea
\Im J_{10}&=&\frac{1}{v^7}\sum_{i,j,k}\epsilon_{ijk}e_i M_i^4  e_j q_k,\\
\Im J_{30}&=&\frac{1}{v^5}\sum_{i,j,k}\epsilon_{ijk} q_i M_i^2  e_j q_k.
\eea
The quantity $\Im J_{10}$ is similar to $\Im J_1$ in the sense that it is bilinear in $e_i$ and linear in $q_i$, whereas $\Im J_{30}$ is linear in $e_i$ and bilinear in $q_i$. 
It is straightforward to show that if both $\Im J_1$ and $\Im J_2$ vanish, then also $\Im J_{10}$ vanishes. Thus, we may conclude the following:\\ \\
{\bf CP is conserved if and only if} {\boldmath $\Im J_1=\Im J_2=\Im J_{30}=0$}.\\ \\
The reason for using $\Im J_{30}$ instead of $\Im J_{3}$ is that  $\Im J_{30}$ is much easier to connect directly to an experimentally observable quantity due to its ``atomic" form.

It is also worth noting that one can write these expressions as determinants:
\begin{align}
\Im J_1 
&=\frac{1}{v^5}
\left|
\begin{matrix}
q_1 & q_2 & q_3 \\
e_1 & e_2 & e_3 \\
e_1M_1^2 & e_2M_2^2 & e_3M_3^2
\end{matrix}
\right|, \\[4mm]
\Im J_2
&=\frac{2}{v^9}
\left|
\begin{matrix}
e_1 & e_2 & e_3 \\
e_1M_1^2 & e_2M_2^2 & e_3M_3^2 \\
e_1M_1^4 & e_2M_2^4 & e_3M_3^4 
\end{matrix}
\right|, 
\\[4mm]
\Im J_{30}
&=\frac{1}{v^5}
\left|
\begin{matrix}
e_1 & e_2 & e_3 \\
q_1 & q_2 & q_3 \\
q_1M_1^2 &q_2M_2^2 &q_3M_3^2
\end{matrix}
\right|.
\end{align}
Comparing these expressions to the invariants found by Lavoura and Silva \cite{Lavoura:1994fv}, who worked in the ``Higgs basis", we see that our expressions reduce to theirs for this particular basis:
\bea
\left[\Im J_2\right]_\text{Higgs-basis} &=&-2v^6 J_1^\text{LS} ,\\
\left[\Im J_1\right]_\text{Higgs-basis} &=&v^3 J_3^\text{LS} ,\\
\left[\Im J_{30}\right]_\text{Higgs-basis} &=&-v^4 J_2^\text{LS} .
\eea
where $J_i^\text{LS}$, $i=1,2,3$ refer to the expressions found by \cite{Lavoura:1994fv}.

%%%%%%%%%%%%%%%%%%%%%%%%%%%%%%%%%%%%%%%%%%%%%%%%%%%%%%%%%%%%%%%%%%%%%%%%%%%%%
\section{An attempt to measure $\Im J_i$}
\label{Sec:INV}
\setcounter{equation}{0}
%%%%%%%%%%%%%%%%%%%%%%%%%%%%%%%%%%%%%%%%%%%%%%%%%%%%%%%%%%%%%%%%%%%%%%%%%%%%%

We here outline a systematic approach to discover or exclude CP violation in the 2HDM, starting with observables which are theoretically easier to interpret. 

%%%%%%%%%%%%%%%%%%%%%%%%%%%%%%%%%%%%%%%%%%%%%%%%%%%%%%%%%%%%%%%%%%%%%%%%%%%%%
\subsection{$\Im J_2$}
\begin{figure}[htb]
%\vspace*{-2.0cm}
\centerline{
\includegraphics[width=5.0cm]{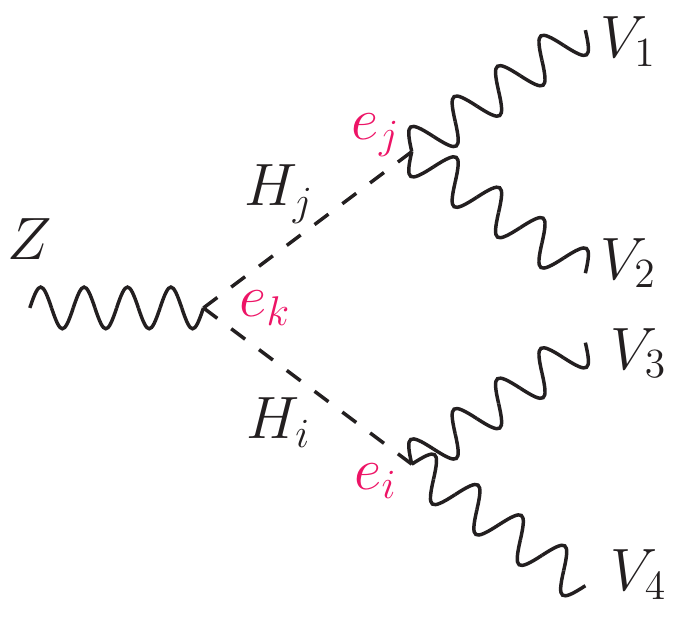}\hspace{2cm}
\includegraphics[width=5.0cm]{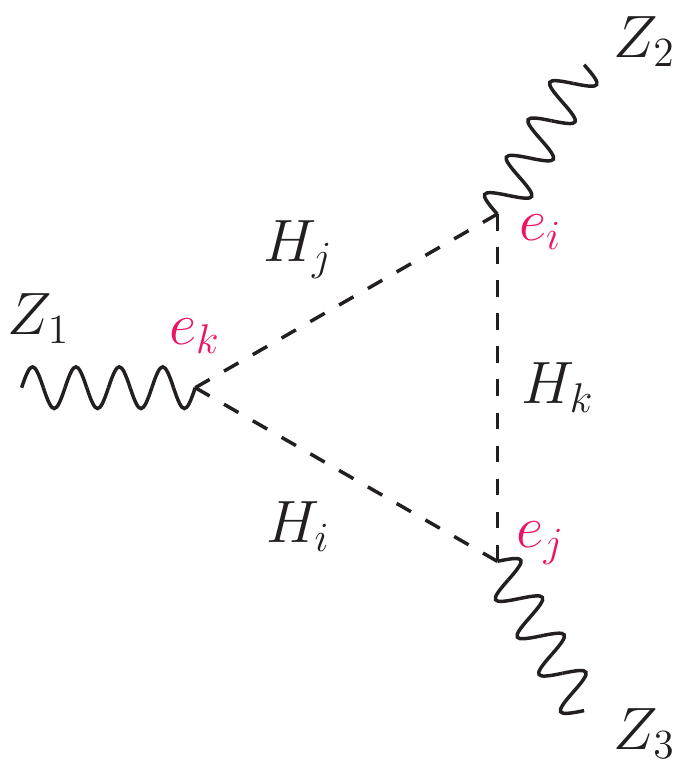}}
\caption{Feynman diagrams related to $\Im J_2$.}
\label{fig:Feynman-j2}
\end{figure}

%%%%%%%%%%%%%%%%%%%%%%%%%%%%%%%%%%%%%%%%%%%%%%%%%%%%%%%%%%%%%%%%%%%%%%%%%%%%%
Since $\Im J_2$ is trilinear in $e_i$ we look for Feynman diagrams containing three vertices, where each vertex is proportional to $e_i$. 
Also, since $\Im J_2$ contains the antisymmetric tensor $\epsilon_{ijk}$, we are led to choose one of the vertices to be $ZH_iH_j$.
This leads us to a study of the Feynman amplitude structures shown in figure \ref{fig:Feynman-j2}.

%%%%%%%%%%%%%%%%%%%%%%%%%%%%%%%%%%%%%%%%%%%%%%%%%%%%%%%%
The amplitude of the left (tree level) diagrams (a total of six diagrams) will be proportional to 
\bea
\mcal\propto \sum_{i,j,k}\epsilon_{ijk}e_ie_je_k\frac{1}{p_{34}^2-M_i^2}\frac{1}{p_{12}^2-M_j^2}
\eea 
where $p_{12}$ is the sum of the momenta of $V_1$ and $V_2$, whereas  $p_{34}$ is the sum of the momenta of $V_3$ and $V_4$. The pairs 
$V_1V_2$ and $V_3V_4$ may be either $ZZ$-pairs or $W^+W^-$-pairs. Performing the sum over all possible combinations of internal $H_i$ and $H_j$ and over $k$, we get
\bea
\mcal\propto\frac{(p_{34}^2-p_{12}^2)v^9\Im J_2}{\displaystyle\prod_{n=1}^3(p_{12}^2-M_n^2)(p_{34}^2-M_n^2)}.
\eea
The amplitude of the right (triangle loop) diagrams (six in total) will be proportional to 
\bea
\mcal  &\propto& \sum_{i,j,k} \int d^4q \sum_{a,b,c}\epsilon_{ija}\epsilon_{kjb}\epsilon_{ikc}e_ae_be_c\frac{1}{q^2-M_i^2}\frac{1}{(q+p_1)^2-M_j^2}\frac{1}{(q+p_1+p_2)^2-M_k^2}\nonumber\\
&&\times(2q+p_1)^{\mu_1}(2q+2p_1+p_2)^{\mu_2}(2q+p_1+p_2)^{\mu_3}
\eea 
where $p_1$ and $p_2$ are the (incoming) momenta of $Z_1$ and $Z_2$, respectively. Performing the sum over $i,j,k$, we find
\bea
\mcal &\propto&\int d^4q\frac{[q^2-(q+p_1)^2][q^2-(q+p_1+p_2)^2][(q+p_1)^2-(q+p_1+p_2)^2]v^9\Im J_2}{\displaystyle\prod_{n=1}^3(q^2-M_n^2)((q+p_1)^2-M_n^2)((q+p_1+p_2)^2-M_n^2)}\nonumber\\
&&\times (2q+p_1)^{\mu_1}(2q+2p_1+p_2)^{\mu_2}(2q+p_1+p_2)^{\mu_3}.
\eea
We see from this that the amplitudes of both these diagrams are directly proportional to $\Im J_2$.

%%%%%%%%%%%%%%%%%%%%%%%%%%%%%%%%%%%%%%%%%%%%%%%%%%%%%%%%%%%%%%%%%%%%%%%%%%%%%
\subsection{$\Im J_1$}
\begin{figure}[htb]
%\vspace*{-2.0cm}
\centerline{
\includegraphics[width=6.0cm]{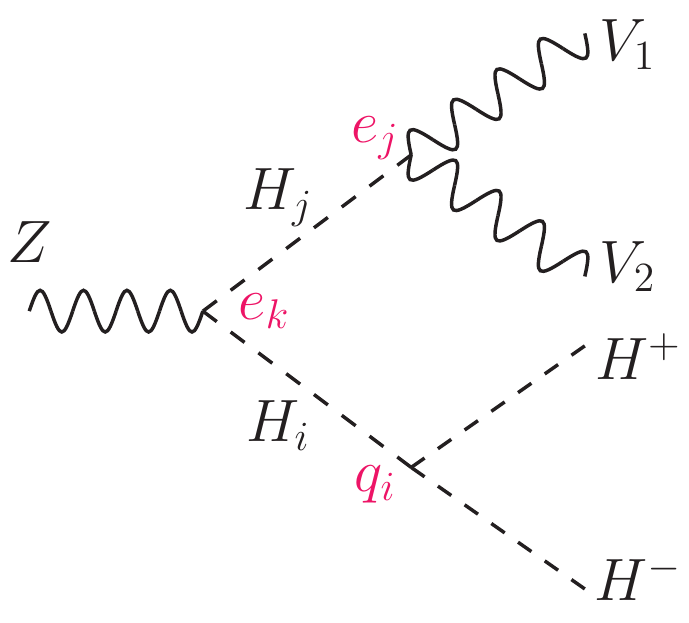}}
\caption{Feynman diagram related to $\Im J_1$.}
\label{fig:Feynman-j1-c2}
\end{figure}

%%%%%%%%%%%%%%%%%%%%%%%%%%%%%%%%%%%%%%%%%%%%%%%%%%%%%%%%%%%%%%%%%%%%%%%%%%%%%
Since $\Im J_1$ is bilinear in $e_i$ and linear in $q_i$, we look for Feynman diagrams containing three vertices, where two of the vertices are proportional to $e_i$, and the third vertex is an $H_iH^+H^-$-vertex. 
Also, since $\Im J_1$ contains the antisymmetric tensor $\epsilon_{ijk}$, we are led to choose one of the vertices to be $ZH_iH_j$. This leads us to a study of the Feynman 
amplitude structure shown in figure \ref{fig:Feynman-j1-c2}.

%%%%%%%%%%%%%%%%%%%%%%%%%%%%%%%%%%%%%%%%%%%%%%%%%%%%%%%%
%%%%%%%%%%%%%%%%%%%%%%%%%%%%%%%%%%%%%%%%%%%%%%%%%%%%%%%%
The amplitude corresponding to the six diagrams shown in figure~\ref{fig:Feynman-j1-c2} is proportional to 
\bea
\mcal \propto \sum_{i,j,k} \epsilon_{ijk}q_ie_je_k\frac{1}{p_{HH}^2-M_i^2}\frac{1}{p_{12}^2-M_j^2}
\eea 
where $p_{12}$ is the sum of the momenta of $V_1$ and $V_2$, whereas  $p_{HH}$ is the sum of the momenta of the $H^+H^-$-pair. The $V_1V_2$-pair may 
be either a $ZZ$-pair or a $W^+W^-$-pair. Summing over $i,j,k$, we get
\bea
\mcal\propto\frac{C_1\Im J_1+C_{11}\Im J_{11}+C_{12}\Im J_{12}}{\displaystyle\prod_{n=1}^3(p_{12}^2-M_n^2)(p_{HH}^2-M_n^2)}
\eea
where
\bea
\Im J_{11}&=&\frac{1}{v^7}\sum_{i,j,k}\epsilon_{ijk}e_i M_i^2 M_j^2 e_k q_j,\\
\Im J_{12}&=&\frac{1}{v^9}\sum_{i,j,k}\epsilon_{ijk}e_i M_i^2 M_j^4 e_k q_j,\\
C_1&=&\left[p_{12}^2 (p_{HH}^2-p_{12}^2) (M_1^2+M_2^2+M_3^2-p_{12}^2-p_{HH}^2)\right.\nonumber\\
&&\left.-(p_{12}^2-M_1^2) (p_{12}^2-M_2^2) (p_{12}^2-M_3^2)\right]v^5, \\
C_{11}&=&\left[p_{HH}^2 - (M_1^2 + M_2^2 + M_3^2)\right](p_{HH}^2-p_{12}^2) v^7, \\
C_{12}&=&(p_{HH}^2-p_{12}^2) v^9.
\eea
The quantities $\Im J_{11}$ and $\Im J_{12}$ are both bilinear in $e_i$ and linear in $q_i$, but have different ``mass weights" compared to $\Im J_1$. Simple algebra shows that they both vanish when $\Im J_1=\Im J_2=0$.

Let us also note the intriguing property that as 
\begin{equation} \label{eq:j1-kin-limit}
p_{HH}^2\to p_{12}^2,
\end{equation}
then $\mcal$ simplifies enormously since $C_{11}=C_{12}=0$, and the total amplitude becomes proportional to $\Im J_1$.
In principle, one could imagine exploiting this property 
experimentally by studying this process for a range of kinematical configurations, and extrapolating to the limit (\ref{eq:j1-kin-limit}).
%%%%%%%%%%%%%%%%%%%%%%%%%%%%%%%%%%%%%%%%%%%%%%%%%%%%%%%%%%%%%%%%%%%%%%%%%%%%%
\subsection{$\Im J_{30}$}
\begin{figure}[htb]
%\vspace*{-2.0cm}
\centerline{
\includegraphics[width=5.0cm]{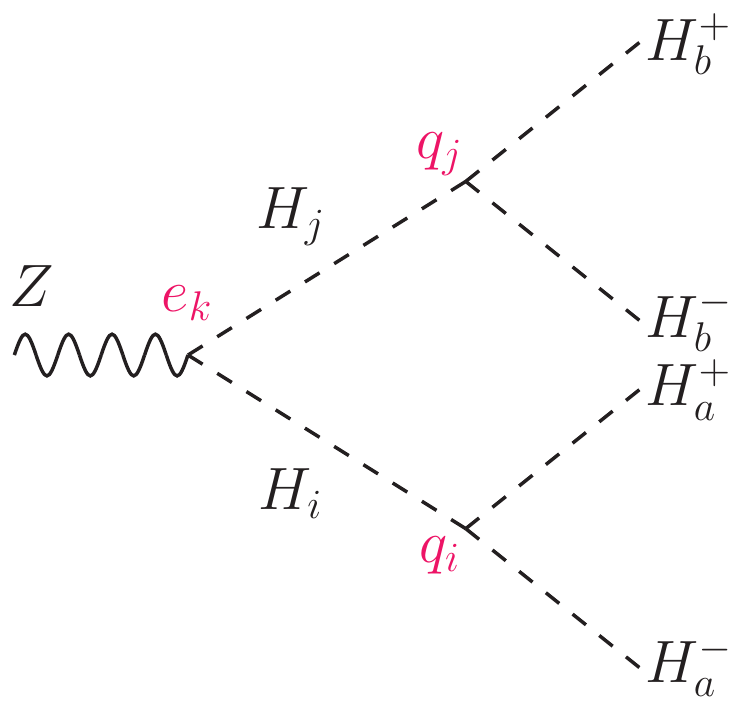}\hspace{2cm}
\includegraphics[width=5.0cm]{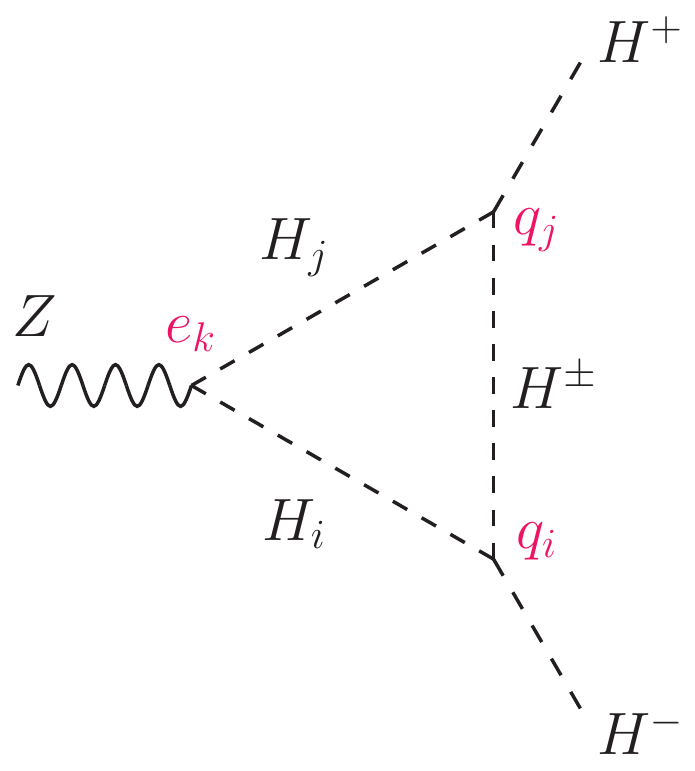}}
\caption{Feynman diagrams related to $\Im J_{30}$.}
\label{fig:Feynman-j3-2}
\end{figure}

%%%%%%%%%%%%%%%%%%%%%%%%%%%%%%%%%%%%%%%%%%%%%%%%%%%%%%%%%%%%%%%%%%%%%%%%%%%%%
Since $\Im J_{30}$ is linear in $e_i$ and bilinear in $q_i$, in this case
we look for Feynman diagrams containing three vertices, where one vertex is proportional to $e_i$, 
and the two other vertices are $H_iH^+H^-$-vertices. 
In order to incorporate $\epsilon_{ijk}$ that is present in $\Im J_{30}$ 
we again choose one of the vertices to be $ZH_iH_j$. This leads us to a study of the Feynman 
amplitude structures shown in figure \ref{fig:Feynman-j3-2}.

%%%%%%%%%%%%%%%%%%%%%%%%%%%%%%%%%%%%%%%%%%%%%%%%%%%%%%%%
%%%%%%%%%%%%%%%%%%%%%%%%%%%%%%%%%%%%%%%%%%%%%%%%%%%%%%%%

The amplitude of the six left (tree level) diagrams will be proportional to 
\bea
\mcal \propto \sum_{i,j,k}\epsilon_{ijk}q_iq_je_k\frac{1}{p_{a}^2-M_i^2}\frac{1}{p_{b}^2-M_j^2}
\eea 
where $p_{a}$ and $p_{b}$ denote the sum of the (outgoing) momenta of the $H_a^-H_a^+$ and $H_b^-H_b^+$ pairs, respectively. Summing over all possible combinations of $i,j,k$, we obtain
\bea
\mcal \propto\frac{(p_a^2-p_b^2)\left[p_a^2p_b^2v^5\Im J_{30}-(p_a^2+p_b^2) v^7\Im J_{31}+v^9\Im J_{32}\right]}{\displaystyle\prod_{n=1}^3(p_{a}^2-M_n^2)(p_{b}^2-M_n^2)}.
\eea
Similarly, the amplitude of the six right (triangle loop) diagrams will be proportional to 
\bea
\mcal  &\propto& \sum_{i,j,k} \int d^4q \epsilon_{ijk}q_iq_je_k \frac{1}{q^2-M_i^2}\frac{1}{(q+p_Z)^2-M_j^2}\frac{1}{(q+p_Z+p_+)^2-M_{H^\pm}^2}\nonumber\\
&&\times(2q+p_Z)^{\mu}
\eea 
where $p_Z$ and $p_+$ are the (incoming) momenta of $Z$ and $H^+$, respectively. Summing over all possible combinations of $i,j,k$ we find
\bea
\mcal
&\propto&\int d^4q\frac{q^2(q+p_Z)^2v^5\Im J_{30}-[q^2+(q+p_Z)^2] v^7\Im J_{31}+v^9\Im J_{32}}{\displaystyle[(q+p_Z+p_+)^2-M_{H^\pm}^2]\prod_{n=1}^3(q^2-M_n^2)((q+p_Z)^2-M_n^2)}\nonumber\\
&&\times [q^2-(q+p_Z)^2] (2q+p_Z)^{\mu}.
\eea
Here,
\bea
\Im J_{31}&=&\frac{1}{v^7}\sum_{i,j,k}\epsilon_{ijk} q_i M_i^2 M_j^2 e_j q_k,\\
\Im J_{32}&=&\frac{1}{v^9}\sum_{i,j,k}\epsilon_{ijk} q_i M_i^2 M_j^4 e_j q_k.
\eea

The quantities $\Im J_{31}$ and $\Im J_{32}$ are both linear in $e_i$ and bilinear in $q_i$, but have different ``mass weights" compared to $\Im J_{30}$. Simple algebra shows that they vanish when $\Im J_1=\Im J_2=\Im J_{30}=0$.

%%%%%%%%%%%%%%%%%%%%%%%%%%%%%%%%%%%%%%%%%%%%%%%%%%%%%%%%%%%%%%%%%%%%%%%%%%%%%
\section{The alignment limit}
\label{Sec:H1SM limit}
\setcounter{equation}{0}
%%%%%%%%%%%%%%%%%%%%%%%%%%%%%%%%%%%%%%%%%%%%%%%%%%%%%%%%%%%%%%%%%%%%%%%%%%%%%

As is well known, the properties of the Higgs boson discovered at the LHC are close to those predicted by the SM~\cite{cms_coupl,Aad:2014eha,Aad:2014tca}.
Motivated by this experimental fact we shall in this section discuss
the limit of the 2HDM which reproduces the SM couplings of $H_1$ to
vector bosons. The limit is referred to as alignment,
see sec.~1.3 in \cite{Asner:2013psa} and
\cite{Craig:2013hca,Carena:2013ooa}.\footnote{``Alignment'' is used also in
  the flavour sector, we emphasize that these are different kinds of ``alignment".}

It should be emphasized that no assumptions concerning the mass spectrum of non-standard Higgs bosons is being made here.
Therefore the alignment limit is not identical to the decoupling limit~\cite{Gunion:2002zf} which is defined by increasing the masses of non-standard Higgs bosons.
Of course, decoupling implies alignment, but the inverse is not true. In fact, it has  recently been verified \cite{Dumont:2014wha} by fitting the 2HDM (type I and II) to available experimental data, that the model indeed allows for masses of extra scalars 
even within the $150-200$~GeV range, which is below the decoupling regime.

Within the CP-violating 2HDM the coupling of $H_1$ to a pair of vector bosons, $e_1$, can be written as:
\beq
e_1 = v \cos(\alpha_2)\cos(\alpha_1-\beta) ,
\label{Eq:e_1}
\eeq
where $\tan\beta=v_2/v_1$. The alignment is equivalent to putting $e_1=v$.
Since the $e_i$ satisfy the unitarity sum rule $\sum_{i=1,2,3}e_i^2=v^2$, alignment implies also $e_2=e_3=0$, meaning 
\beq \label{Eq:h1sm-limit}
\alpha_1=\beta, \quad \alpha_2=0.
\eeq
The rotation matrix in this case becomes
\begin{equation} 
R=
\begin{pmatrix}
R_{11}    &  R_{12}   & R_{13}   \\
R_{21}    &  R_{22}   & R_{23}   \\
R_{31}    &  R_{32}   & R_{33}  
\end{pmatrix}
=
\begin{pmatrix}
c_\beta & s_\beta & 0 \\
-  s_\beta\,c_3 & c_\beta\,c_3  & s_3 \\
s_\beta\,s_3 & - c_\beta\,s_3  & c_3
\end{pmatrix}.
\label{R_lim}
\end{equation}
Note that the mixing matrix could be written in this case as  
\begin{equation} 
R= R_3 R_1 =
\begin{pmatrix}
1 & 0 & 0 \\
0 & c_3  & s_3 \\
0 & -s_3 & c_3
\end{pmatrix}
\begin{pmatrix}
c_\beta & s_\beta & 0 \\
-  s_\beta\ & c_\beta  & 0 \\
0 & 0 & 1
\end{pmatrix}.
\end{equation}
The couplings between $H_i$ and $H^+H^-$ simplify in the alignment limit:\footnote{Here we adopt a weak basis such that the relative phase of the two vevs vanishes, i.e. $\xi=0$.}
\begin{eqnarray}
q_{1}
&=&\frac{1}{v}\left(2 M_{H^\pm}^2-2\mu^2+M_1^2\right),
\label{q1}\\
q_{2}
&=&
+c_3\left[\frac{(\cb^2-\sb^2)}{v \cb \sb}(M_2^2-\mu^2)
+\frac{v}{2 \sb^2 }\Re\lambda_6
-\frac{v}{2 \cb^2}\Re\lambda_7\right]
+s_3\frac{v}{2 \cb \sb}\Im\lambda_5,
\label{q2}\\
q_{3}
&=&
-s_3\left[\frac{(\cb^2-\sb^2)}{v \cb \sb}(M_3^2-\mu^2)
+\frac{v}{2 \sb^2}\Re\lambda_6
-\frac{v}{2 \cb^2}\Re\lambda_7\right]
+c_3\frac{v}{2 \cb \sb}\Im\lambda_5.
\label{q3}
\end{eqnarray}
It is easy to see that in this limit, the expressions for the CP-violating invariants become\footnote{When $\alpha_2=0$, but $\alpha_1\neq\beta$, we confirm that $\Im J_2$ has the form given in footnote [21] of \cite{Barroso:2012wz} (denoted $J_1$ by them).} 
\begin{eqnarray}
\Im J_1 &=&0,\nonumber\\
\Im J_2 &=&0,\\
\Im J_{30} &=& \frac{q_2 q_3}{v^4}(M_3^2-M_2^2).\nonumber
\end{eqnarray}
Two comments are here in order. First, note that $e_1=v$ implies no CP violation in the couplings to gauge bosons ($\Im J_2 = 0$), 
the only possible CP violation may appear in cubic scalar  couplings $(H_2H^+H^-)$ 
and $(H_3H^+H^-)$, proportional to $q_2$ and $q_3$, respectively.
Second, the necessary condition for CP violation is that both $(H_2 H^+H^-)$ {\it and} $(H_3 H^+H^-)$ couplings must exist
{\it together} with a non-zero $Z H_2H_3$ vertex ($\propto e_1$).  
Note that the existence of the latter implies that for CP invariance, either
$H_2$ or $H_3$ would have to be odd under CP. However, since they {\it both} couple to $H^+H^-$ (that
is CP even), there would be no way to preserve CP.

It is important to note that in the case when $\lambda_6=\lambda_7=0$ (due to $\zBB_2$ symmetry imposed on the dimension-4 part of the potential)
the $(1,3)$ and $(2,3)$ entries of the neutral mass-squared matrix,
${\cal M}^2_{13}$ and ${\cal M}^2_{23}$, are related as follows
\begin{equation} 
{\cal M}^2_{13} = t_\beta {\cal M}^2_{23},
\label{corel}
\end{equation}
where $t_\beta\equiv \tan\beta$. 
As a consequence of the above relation there is a constraint that relates mass eigenvalues, mixing angles and $t_\beta$~\cite{Khater:2003wq}:
\begin{equation} 
M_1^2 R_{13}(R_{12}t_\beta-R_{11}) + M_2^2 R_{23}(R_{22} t_\beta-R_{21}) + M_3^2 R_{33}(R_{32}t_\beta-R_{31}) = 0.
\end{equation}
In the alignment limit, the above relation simplifies to
$(M_2^2-M_3^2) s_3 c_3 s_\beta = 0$, so that either $M_2=M_3$, $\alpha_3=0$ or $\alpha_3=\pm\pi/2$. It is easy to see that in all three cases,
CP violation disappears. If $M_2=M_3$ then after the following reparametrization of $H_i$
\begin{equation} 
\begin{pmatrix}
H_1 \\ H_2 \\ H_3
\end{pmatrix}
\to R_3
\begin{pmatrix}
H_1 \\ H_2 \\ H_3
\end{pmatrix},
\label{rot3}
\end{equation}
the resulting mixing matrix is just $R=R_1$, implying no CP violation, consistent with $\Im J_{30} = 0$. 
If, on the other hand, $\alpha_3=0$ or $\pm\pi/2$, then $q_2=0$ or $q_3=0$, respectively, so again $\Im J_{30}=0.$\footnote{In both
cases the mixing matrix reduces to just $R=R_1$ as in the CP-conserving 2HDM.} It must be emphasized that the above important
conclusion was based on the assumption that $\lambda_6=\lambda_7=0$.

If, on the other hand, $\lambda_6 \neq 0$ and/or $\lambda_7\neq 0$, then ${\cal M}^2_{13}$ and ${\cal M}^2_{23}$
are not correlated as in (\ref{corel}), and we may not claim that $M_2=M_3$, nor that $\alpha_3=0$ or $\pm\pi/2$. So there is still room for CP violation.

Attempts to find symmetries that would naturally lead to alignment severely restrict the model.
One possibility\footnote{We thank Howard Haber for a discussion concerning this point.}
is just the standard $\zBB_2$ (invoked usually upon dimension-4 terms to suppress FCNC in Yukawa couplings) imposed in the Higgs basis, see sec.~1.3 in \cite{Asner:2013psa}.
This symmetry is however much too restrictive as it implies both $\lambda_6=0$ and $\lambda_7=0$ (possibly together with $m_{12}^2=0$),
while the alignment comprises just one constraint i.e., $e_1=v$.
Obviously, when the symmetry
is imposed there is no way to accommodate CP violation in the scalar potential.
Another attempt to find alignment was discussed recently in \cite{Dev:2014yca}, where the authors introduce a 2HDM
based on the SO(5) group and show that this leads to alignment. 
Dimension-4 terms in the scalar potential are assumed to be invariant under SO(5), while the symmetry
is softly broken by bilinear Higgs mass terms. In addition the symmetry is violated by the hypercharge gauge coupling 
and third-generation Yukawa couplings.
Again, the symmetry is so restrictive that there is no room for CP-violation in the scalar potential within this scenario.

So far, we have defined alignment in terms of mixing angles, $\alpha_1=\beta$ and $\alpha_2=0$. However, it is also worth trying to express the conditions in terms of the potential parameters. 
Especially, in the case of seeking a symmetry responsible for the alignment 
it is necessary to have the alignment condition in terms of scalar quartic 
coupling constants. Let us start with the relation between the initial, non-diagonal
scalar mass-squared matrix ${\cal M}^2$ and the mixing angles. In general we have
\begin{equation}
R{\cal M}^2R^{\rm T}={\cal M}^2_{\rm diag}={\rm diag}(M_1^2,M_2^2,M_3^2).
\nonumber
\end{equation}
In the case of alignment, $R=R_3 R_\beta$ where $R_\beta\equiv R_1|_{\alpha_1=\beta}$.
Then the mass-squared matrix ${\cal M}^2$ must be diagonalizable by $R_3 R_\beta$.
Therefore it is of the following form:
\beq
{\cal M}^2 = R_\beta^T R_3^T {\cal M}^2_{\rm diag} R_3 R_\beta.
\label{alig_gen}
\eeq
For a given $t_\beta$ the above form of ${\cal M}^2$ has 4 independent parameters, 
while in general it would have 6 parameters. Therefore we anticipate the existence
of two relations between the a priori independent entries of ${\cal M}^2$. 
Those relations would be the wanted alignment conditions, expressed in terms 
of the potential parameters
and $t_\beta$. Indeed, from (\ref{alig_gen}) one can find the following constraints 
satisfied by the entries of ${\cal M}^2$:
\bea
{\cal M}_{13}^2 &=&-t_\beta {\cal M}_{23}^2, \label{alig_con1}\\
\left(t_\beta^{-1} -t_\beta\right) {\cal M}_{12}^2 &=& ({\cal M}_{11}^2-{\cal M}_{22}^2). \label{alig_con2}
\eea
It is worth noting that these two formulas are satisfied not only by $e_1=v$,  in fact they hold 
whenever $e_i^2=v^2$ for any $i=1,2,3$. 
Using the general formulae for the elements of ${\cal M}^2$, (\ref{M11})--(\ref{M23}) one finds
from (\ref{alig_con1})--(\ref{alig_con2})
\bea
v_2^2 \Im\left(e^{i\xi}\lam_7 \right)+v_2 v_1 \Im\left(e^{2i\xi}\lam_5 \right)               
+v_1^2 \Im\left(e^{i\xi}\lam_6 \right) &=& 0 , \label{alig_lam1}\\     
v_2^4  \Re\left(e^{i\xi}\lam_7\right)  
+v_2^3v_1(-\lambda_2+\lam_{345})
+3v_2^2v_1^2 \Re\left[e^{i\xi}(\lam_6-\lam_7)\right]\nonumber + && \\
+v_2v_1^3(\lambda_1-\lam_{345})
-v_1^4  \Re\left(e^{i\xi}\lam_6\right) &=& 0
\label{alig_lam2}
\eea
where $\lam_{345}\equiv \lambda_3+\lambda_4+\Re\left(e^{2i\xi}\lam_5\right)$,
and the terms have been ordered in powers of $v_1$ and $v_2$.
In the CP-conserving limit, with $\xi=0$, $\Im\lambda_5=\Im\lambda_6=\Im\lambda_7=0$, 
we reproduce the single alignment condition found recently in ref.~\cite{Dev:2014yca}.
If one wishes to satisfy the alignment conditions for any value of $v_1$, $v_2$ and $\xi$, then 
the following constraints must be fulfilled:
\beq
\lam_1=\lam_2=\lam_3+\lam_4, \;\; \lam_5=\lam_6=\lam_7=0.
\label{gen_align}
\eeq
Two comments are here in order. First, the above constraints eliminate the possibility of CP violation. 
Seeking a relation between quartic coupling constants that would be
responsible for the alignment one would indeed have to satisfy (\ref{alig_lam1})--(\ref{alig_lam2}) without any reference
to the $v_1$, $v_2$ and $\xi$. Unfortunately its implication (\ref{gen_align}) is  inconsistent with 
the possibility of having CP violated in the potential.
However, the reader should be reminded that from a phenomenological point of view there is 
no need to satisfy the alignment conditions regardless of the value of $v_2/v_1$. 
We may conclude that for CP to be violated, $\tb$ must be properly tuned
to satisfy the alignment conditions. Second, it is amusing to note that a potential
satisfying the conditions (\ref{gen_align}) in a CP-conserving case,
has been considered in \cite{Chankowski:2000an} in a different context, namely that of finding a 2HDM that would automatically satisfy the $S,T,U$ constraints. 
An underlying symmetry has not been determined.

We can conclude that the observation of the SM-like Higgs boson at the LHC 
implies (within the 2HDM with $\zBB_2$ softly broken) vanishing CP violation in the scalar potential. 
Note that this conclusion could be realized either by large masses of the extra Higgs bosons 
(the decoupling regime, the case we are not discussing)
or by alignment with relatively light extra Higgs bosons (the case discussed here). 
For both possibilities the $H_1VV$ coupling is SM-like 
so CP violation disappears within the 2HDM with softly broken $\zBB_2$ symmetry. 
We can summarize by emphasizing that, in order for CP violation to be present in the scalar potential, then the LHC data would favour a generic 2HDM with no $\zBB_2$ symmetry (thus allowing for non-zero  $\lambda_6$ and/or $\lambda_7$).
A consequence of that would be the interesting  possibility of large (tree-level generated) FCNC in some Yukawa couplings.

%%%%%%%%%%%%%%%%%%%%%%%%%%%%%%%%%%%%%%%%%%%%%%%%%%%%%%%%%%%%%%%%%%%%%%%%%%%%%
\section{Numerical illustrations}
\label{Sec:Illustrations}
\setcounter{equation}{0}
%%%%%%%%%%%%%%%%%%%%%%%%%%%%%%%%%%%%%%%%%%%%%%%%%%%%%%%%%%%%%%%%%%%%%%%%%%%%%

In reality measurements will never tell us that indeed the $H_1VV$ coupling is exactly SM-like, so we should allow for some maximal deviation from alignment. For that purpose we define 
\begin{equation} \label{Eq:delta}
\delta\equiv 1- e_1/v 
\end{equation}
and illustrate predictions for the remaining CP violation as functions of $\delta$. 

First we specify two cases of our scanning strategy. 
\bi
\item 
We denote by 2HDM5 the model with $\zBB_2$ softly broken, so $\lam_6=\lam_7=0$. The model parameters are listed as 
$$
\pcal_5\equiv\{M_{H^\pm}^2,\mu^2,M_1^2,M_2^2, v_1, v_2, \xi=0,\alpha_1,\alpha_2,\alpha_3\}.
$$
In this case we fix $M_{H^\pm}^2$, $\mu^2$, $M_2$, $\tan\beta$  (for the LHC Higgs boson we use $M_1=125\gev$) 
and scan over $-\pi/2 \leq \alpha_1,\alpha_2,\alpha_3 \leq \pi/2$ for chosen maximal deviation $\delta$,
imposing $M_1<M_2<M_3$, vacuum stability and unitarity.
\item 
2HDM67 refers to the general 2HDM, so that $\zBB_2$ is not imposed, consequently $\lam_6\neq 0, \lam_7\neq 0$.
In this case the parameters of the model are
$$
\pcal_{67}\equiv\{M_{H^\pm}^2,\mu^2,M_1^2,M_2^2,M_3^2,{\rm Im}\lambda_5,{\rm Re}\lambda_6,{\rm Re}\lambda_7, v_1, v_2, \xi=0,\alpha_1,\alpha_2,\alpha_3\}
$$
For this general case $M_{H^\pm}^2$, $\mu^2$, $M_2$, $M_3$, and $\tan\beta$  are fixed (for the LHC Higgs boson we use $M_1=125\gev$)  while 
we scan over $-\pi/2\leq \alpha_1,\alpha_2,\alpha_3\leq \pi/2$ and $|{\rm Im}\lambda_5|,|{\rm Re}\lambda_6|,|{\rm Re}\lambda_7|<5$,
for chosen maximal deviation $\delta$, imposing $M_1<M_2<M_3$, vacuum stability and unitarity.
\ei

\begin{figure}[htb]
\centering
\includegraphics[width=16 cm,height=8 cm]{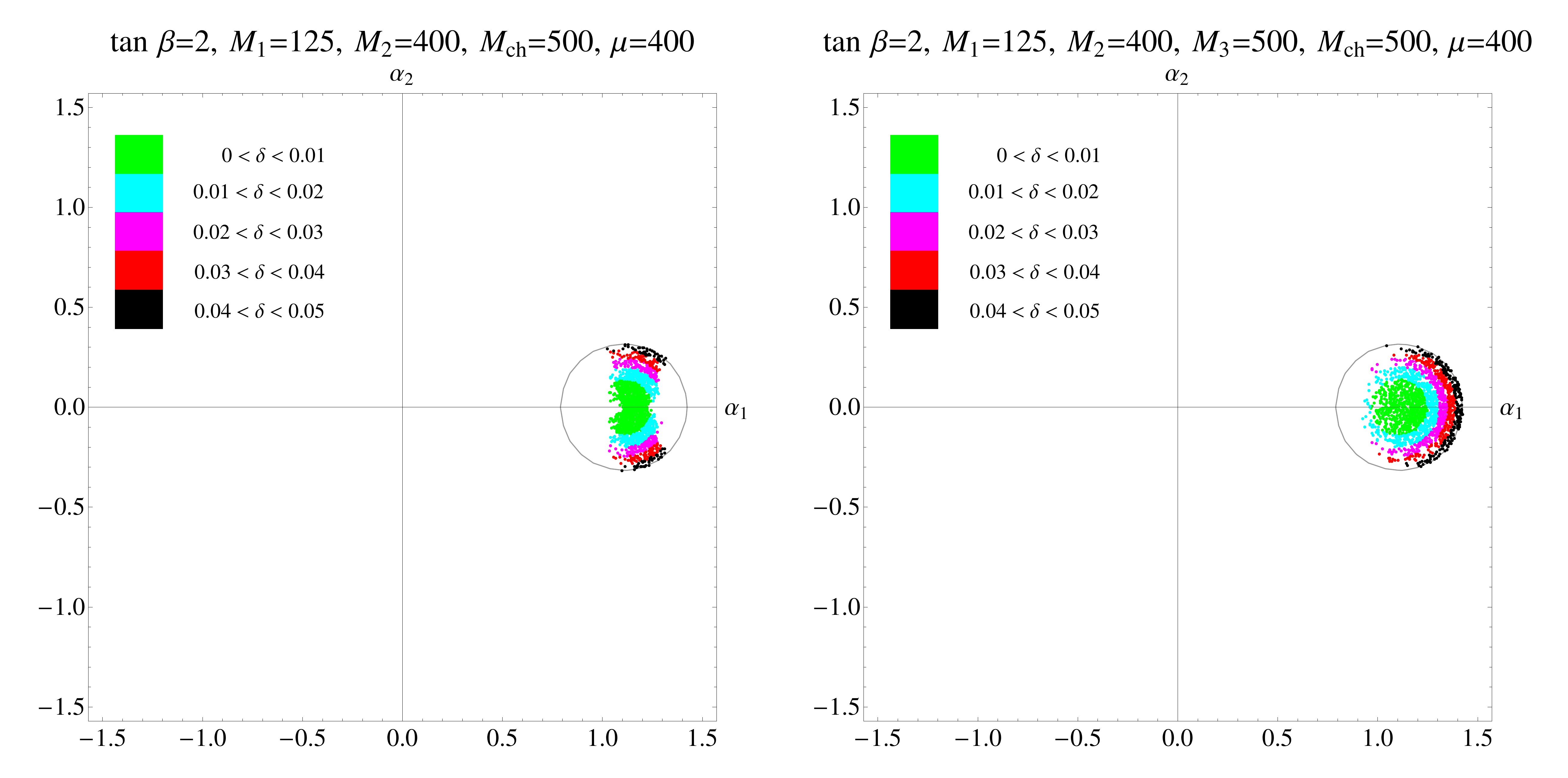}
\caption{Allowed regions in the $(\alpha_1,\alpha_2)$ space for $\tan\beta=2$ corresponding to
maximal deviation $\delta=0.05$ within the 2HDM5 ($\zBB_2$ softly broken) and 2HDM67, are shown in the left and right panels, respectively. 
The coloring corresponds to ranges of $\delta$ shown in the legend. 
Vacuum stability and unitarity constraints are satisfied. 
The parameters adopted are $M_1=125$ GeV, $M_2=400$ GeV, $M_{H^\pm}=500$ GeV, $\tan\beta=2$, $\mu=400$ GeV for the 2HDM5. For the 2HDM67, in addition, $M_3=500$ GeV.}
\label{fig:delta}
\end{figure}

Figure~\ref{fig:delta} illustrates regions of $(\alpha_1,\alpha_2)$ (see Eq.~(\ref{Eq:h1sm-limit})) which are compatible with $\delta \leq 0.05$, the external edge corresponds to $\delta=0.05$. Approaching the center of the contours $\delta \to 0$.

\begin{figure}
\centering
\includegraphics[width=16cm, height=8 cm]{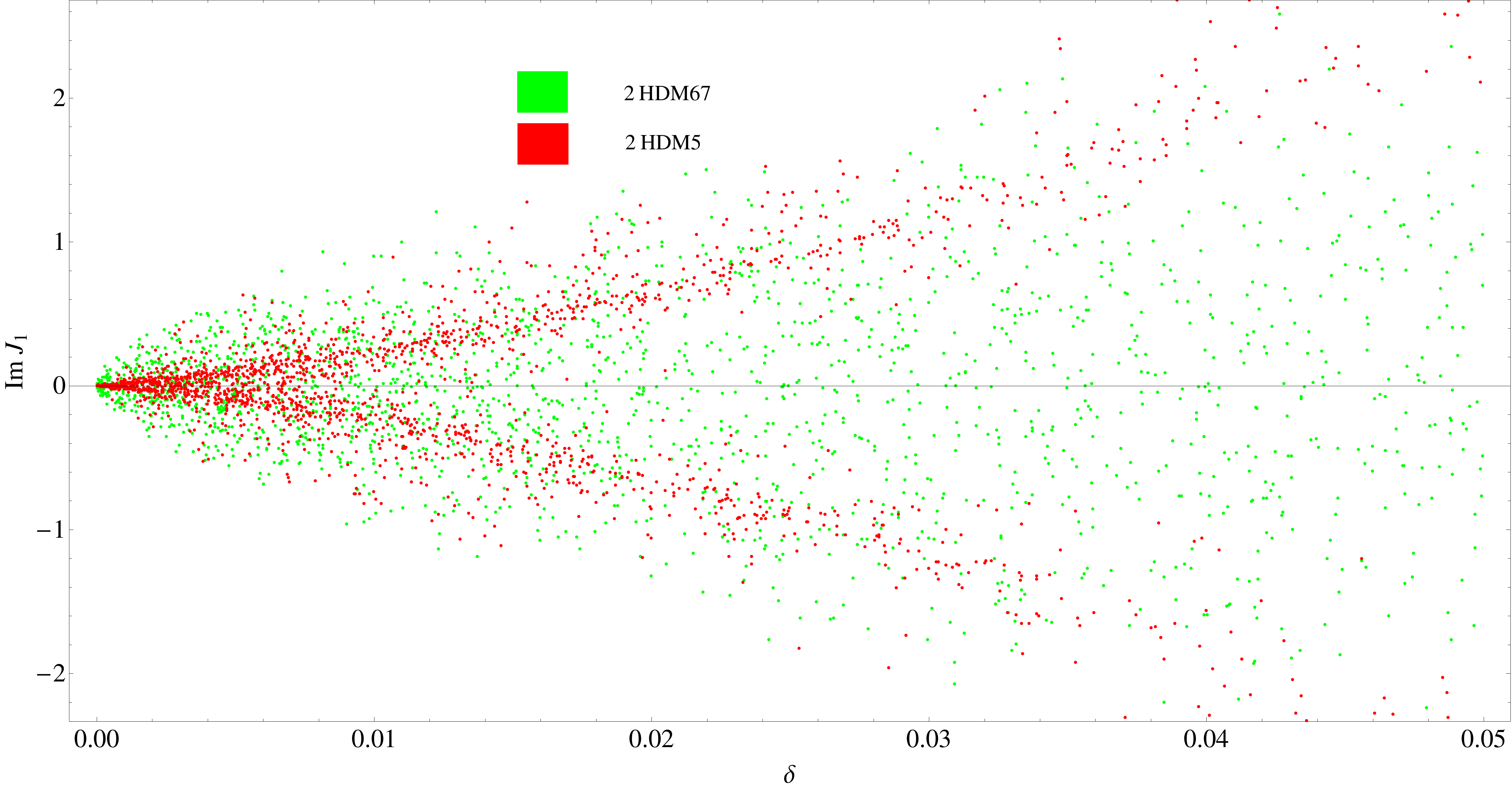}
\caption{Correlation between $\Im J_1$ and the deviation $\delta$ for $\delta\leq 0.05$. 
Green and red dots correspond to 2HDM67 and 2HDM5, respectively.
}
\label{deltavsImJ1}
\end{figure}

\begin{figure}
\centering
\includegraphics[width=16 cm,height=8 cm]{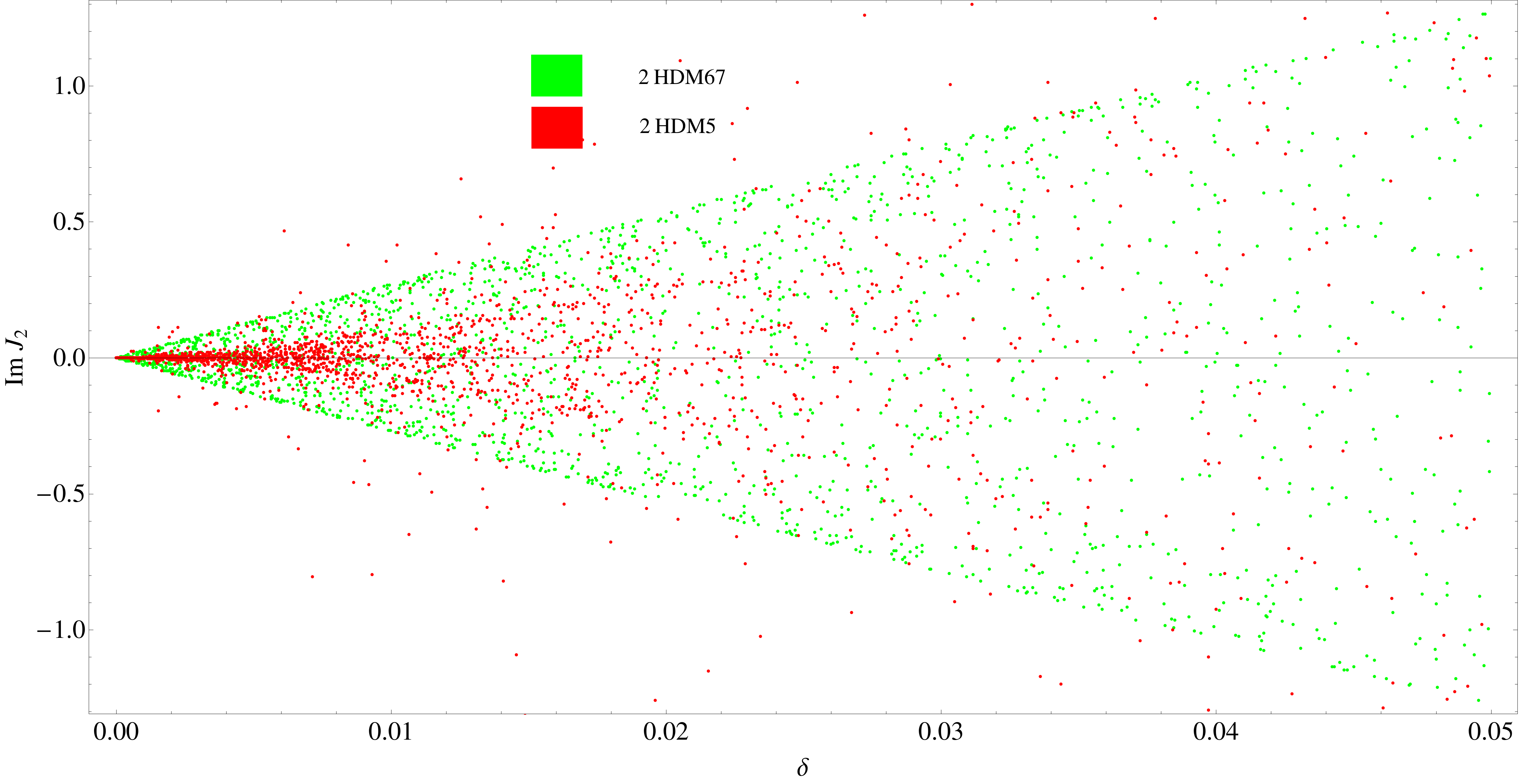}
\caption{Correlation between $\Im J_2$ and the deviation $\delta$ for $\delta\leq 0.05$. 
Green and red dots correspond to 2HDM67 and 2HDM5, respectively.
}
\label{deltavsImJ2}
\end{figure}

\begin{figure}
\centering
\includegraphics[width=16 cm,height=8 cm]{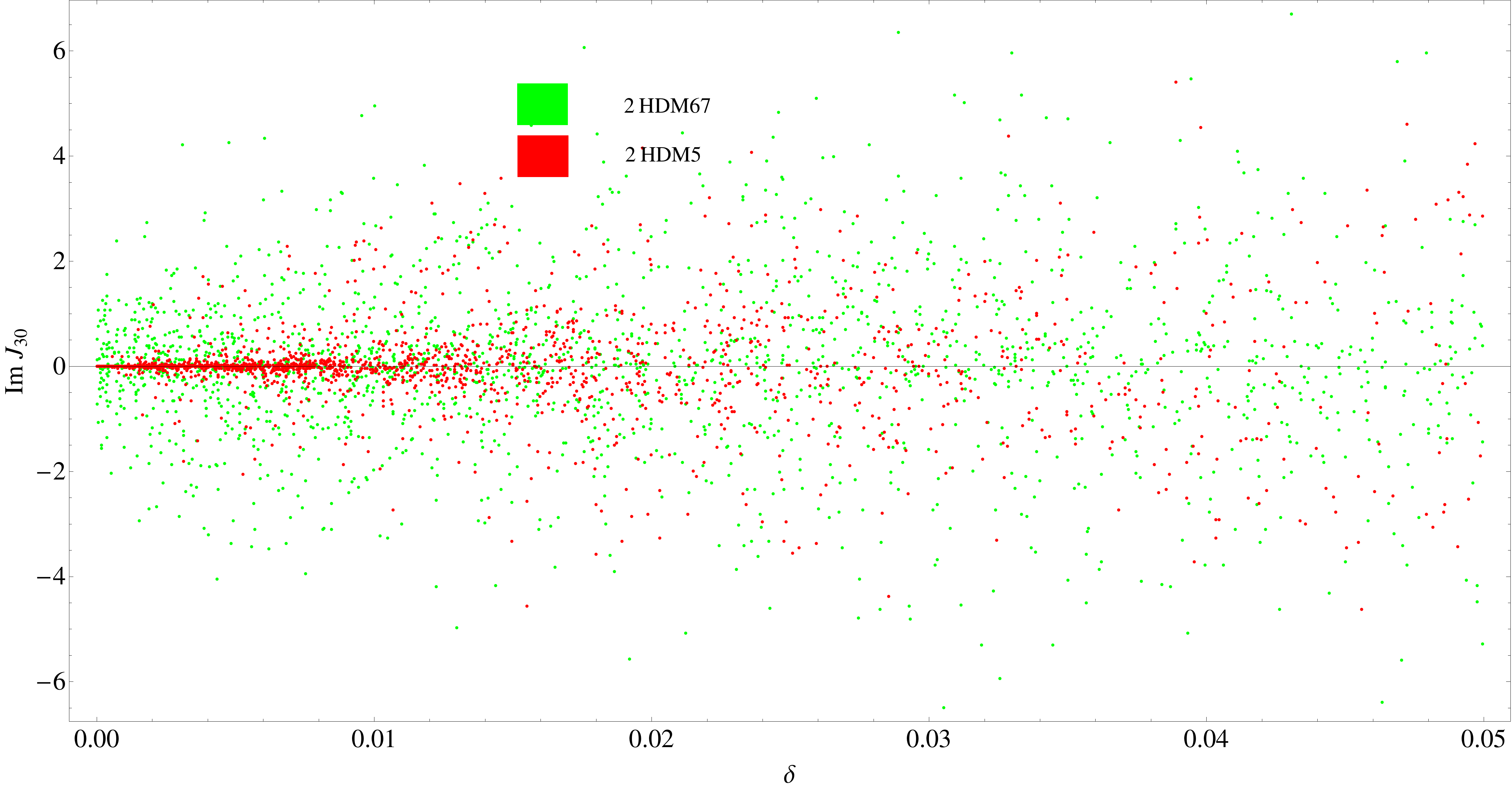}
\caption{Correlation between $\Im J_{30}$ and the deviation $\delta$ for $\delta\leq 0.05$. 
Green and red dots correspond to 2HDM67 and 2HDM5, respectively.
}
\label{deltavsImJ30}
\end{figure}

In figures~\ref{deltavsImJ1}--\ref{deltavsImJ30} we show correlations between $\Im J_i$ and $\delta$.
As one could have anticipated from the discussion of alignment, all the invariants
must vanish as $\delta$ decreases in the model with $\lam_6=\lam_7=0$. However,
as seen in Fig.~\ref{deltavsImJ30}, when $\lam_6\neq 0$ and/or $\lam_7\neq 0$ then
$\Im J_{30}$ does not vanish even when $\delta\to 0$, as illustrated by the green dots for small values of $\delta$, corresponding to non-zero $\Im J_{30}$. Typically $\Im J_{30} \sim 1-3$, showing a large amount of CP violation present in the model even in the vicinity of the alignment limit.

Before closing this section let us focus on the figure showing $\Im J_2$, where remarkably the green dots are all 
inside a triangular region with a clear boundary. In order to understand this, we recall that 
\bea
\Im J_2&=&\frac{2e_1 e_2 e_3}{v^9}(M_1^2-M_2^2)(M_2^2-M_3^2)(M_3^2-M_1^2)\nonumber
\eea
Using the fact that $e_1=v(1-\delta)$ and $|e_3|=\sqrt{v^2-e_1^2-e_2^2}$, one can easily find
that for small $\delta$, $|\Im J_2|_{\rm max}$ simplifies to a linear function reproducing the triangle shape
 \bea
|\Im J_2|_{\rm max}&\simeq&\frac{\delta }{v^6}(M_1^2-M_2^2)(M_2^2-M_3^2)(M_3^2-M_1^2).
\eea

%%%%%%%%%%%%%%%%%%%%%%%%%%%%%%%%%%%%%%%%%%%%%%%%%%%%%%%%
%%%%%%%%%%%%%%%%%%%%%%%%%%%%%%%%%%%%%%%%%%%%%%%%%%%%%%%%

%%%%%%%%%%%%%%%%%%%%%%%%%%%%%%%%%%%%%%%%%%%%%%%%%%%%%%%%
%%%%%%%%%%%%%%%%%%%%%%%%%%%%%%%%%%%%%%%%%%%%%%%%%%%%%%%%

%%%%%%%%%%%%%%%%%%%%%%%%%%%%%%%%%%%%%%%%%%%%%%%%%%%%%%%%
%%%%%%%%%%%%%%%%%%%%%%%%%%%%%%%%%%%%%%%%%%%%%%%%%%%%%%%%

%%%%%%%%%%%%%%%%%%%%%%%%%%%%%%%%%%%%%%%%%%%%%%%%%%%%%%%%
%%%%%%%%%%%%%%%%%%%%%%%%%%%%%%%%%%%%%%%%%%%%%%%%%%%%%%%%

%%%%%%%%%%%%%%%%%%%%%%%%%%%%%%%%%%%%%%%%%%%%%%%%%%%%%%%%%%%%%%%%%%%%%%%%%%%%%
\section{Search strategy}
\label{Sec:Strategy}
\setcounter{equation}{0}
%%%%%%%%%%%%%%%%%%%%%%%%%%%%%%%%%%%%%%%%%%%%%%%%%%%%%%%%%%%%%%%%%%%%%%%%%%%%%
We have identified processes related to all the $\Im J_i$, so that now we are able to propose a strategy 
to test if all the $\Im J_i$ vanish (implying CP conservation), or if one of them is nonzero (CP violation).
Since in the alignment limit only $\Im J_{30}$ might be non-vanishing we focus on prospects for 
its measurement.

\noindent
{\bf Step 1:} Let us start with $\Im J_2$, and choose one of the processes shown in figure \ref{fig:Feynman-j2}. 
The triangle-loop diagram to the right is more appealing due to the fact that the 
$ZZZ$-vertex is absent at tree-level. As we have shown, the amplitude is directly proportional to $\Im J_2$, 
and is thus suitable to determine whether $\Im J_2$ vanishes or not. 

Phenomenological discussions \cite{Hagiwara:1986vm,Gounaris:2000dn,Baur:2000ae} of the $ZZZ$ vertex have presented its most general Lorentz structure.
In Ref.~\cite{Gounaris:2000dn} the CP-violating vertex is analyzed, with $Z_1, Z_2, Z_3$ all off-shell.
A total of 14 Lorentz structures are identified, all preserving parity. Some of these vanish when one or more $Z$ is on-shell. If we characterize them by momenta and Lorentz indices ($p_1,\mu_1$), ($p_2,\mu_2$) and ($p_3,\mu_3$), and let $Z_2$ and $Z_3$ be on-shell, then the structure reduces to the form 
\begin{equation} \label{eq:f_4Z}
e\Gamma_{ZZZ}^{\mu_1\mu_2\mu_3}
=ief_4^Z\frac{p_1^2-M_Z^2}{M_Z^2}(p_1^{\mu_2}g^{\mu_1\mu_3}+p_1^{\mu_3}g^{\mu_1\mu_2}),
\end{equation}
where $e$ is the proton charge and $f_4^Z$ is a dimensionless form factor.
This structure arises from an effective CP-violating operator of the form \cite{Gounaris2:2000}
\bea
{\cal O}_{ZZZ}=\frac{-e}{m_Z^2}f_4^Z\left[\partial_{\mu}\left(\partial^{\mu} Z^{\beta}-\partial^{\beta} Z^{\mu}\right)\right]Z_{\alpha} \left(\partial^{\alpha} Z_{\beta}\right).
\eea
A more detailed phenomenological discussion of $f_4^Z$ will be presented elsewhere \cite{GOO}.

\noindent
{\bf Step 2:}
If in Step 1 we have been able to determine a nonzero $\Im J_2$, then we know that the 2HDM violates CP. If, on the other hand,
we find that $\Im J_2$ is consistent with zero, then the next step would be to proceed to $\Im J_1$, in order to determine if this quantity vanishes or not. For the process shown in figure \ref{fig:Feynman-j1-c2}, we have already shown that in the case of $\Im J_2=0$, the amplitude will vanish if $\Im J_1$ vanishes. We choose $V_1V_2$ to be a $W^+W^-$-pair since it makes all particles involved distinguishable. Characterizing $Z, W^-$ and $W^+$ by Lorentz indices $\mu_1$, $\mu_2$ and $\mu_3$, respectively, the Lorentz-structure for this process is proportional to 
\bea
(p_{VV}-p_{HH})^{\mu_1} g^{\mu_2 \mu_3}=(p_{Z}-2p_{HH})^{\mu_1} g^{\mu_2 \mu_3}.
\eea 

The effective Lagrangian giving rise to such a structure would be
\bea
\left[(\partial_\mu Z^\mu)H^+H^--2Z^\mu(H^-\partial_\mu H^++H^+\partial_\mu H^-)
\right]W_\alpha^\dagger W^\alpha,
\eea
which also clearly violates CP. Again, it is beyond the scope of this work to discuss further phenomenological consequences
of this amplitude. \\
{\bf Step 3:}
If we by now have confirmed that both  $\Im J_1$ and  $\Im J_2$ vanish, the only possibility for a CP-violating 2HDM would be if  $\Im J_{30}$ were nonzero, so we turn to the processes in figure \ref{fig:Feynman-j3-2}. Of these processes, the triangle-loop diagram is more appealing due to fewer particles in the ``final" state. 
The Lorentz structure for this process will consist of two parts:
\bea
Ap_Z^{\mu}+B(p_+-p_-)^{\mu}.
\eea
Here, $A$ will be a CP-violating form factor, and $B$ will be a CP-conserving form factor.
The effective Lagrangian giving rise to the CP-violating part would be
\bea
{\cal O}_{ZH^+H^-}^{CPV}=Z_\mu \left[
(\partial^\mu H^+)H^- +(\partial^\mu H^-)H^+
\right],
\eea
and for the CP conserving one
\bea
{\cal O}_{ZH^+H^-}^{CPC}=Z_\mu \left[
(\partial^\mu H^+)H^- -(\partial^\mu H^-)H^+
\right].
\eea
At the tree-level, the contribution to the $ZH^+H^-$-vertex is proportional to $(p_+-p_-)^{\mu}$, which is CP-conserving. 
CP-violating contributions (proportional to $p_Z^{\mu}$) arise only at loop level. 
In experiment, we would need to single out the parts proportional to $p_Z^{\mu}$ 
in order to measure whether $\Im J_{30}$ vanishes or not. 
In the SM, $A$ (arising from the CKM matrix) would also be generated via quark loops, however a non vanishing 
contribution requires at least a 3-loop diagram, so it is efficiently suppressed.

%%%%%%%%%%%%%%%%%%%%%%%%%%%%%%%%%%%%%%%%%%%%%%%%%%%%%%%%%%%%%%%%%%%%%%%%%%%%%
\section{Summary}
\label{Sec:Summary}
\setcounter{equation}{0}
%%%%%%%%%%%%%%%%%%%%%%%%%%%%%%%%%%%%%%%%%%%%%%%%%%%%%%%%%%%%%%%%%%%%%%%%%%%%%
In this paper we have expressed the three invariants $\Im J_i$ in terms of physically observable quantities like couplings and masses, independently confirming the result of \cite{Lavoura:1994fv,Botella:1994cs}. We have used this to formulate conditions for CP conservation in terms of couplings and masses. We have also identified processes that are sensitive to each of the $\Im J_i$, in order to figure out how to determine, through experiment, whether the 2HDM is CP violating or not.

We have further investigated the scenario in which the lightest neutral scalar mimics the SM Higgs boson, showing that there is still room for CP violation, provided we have nonzero $\lambda_6$ and/or $\lambda_7$, i.e., non-conserved $\zBB_2$ symmetry.

We have also found that there is a conflict between the possibility of violating CP in the potential 
and explanation of the alignment by relations (perhaps symmetry) satisfied by the
quartic coupling constants in the scalar potential for any $v_1$, $v_2$ and the relative phase $\xi$.

Bearing in mind the Higgs LHC data, we have sketched a strategy that may lead to an experimental test of CP violation in the scalar sector of the 2HDM. 
A detailed experimental study of $ZZZ$ and $ZH^+H^-$ vertices together with $Z^\star\to VV H^+H^-$ seem to be necessary in order to test CP symmetry in the scalar potential. 
The proposed strategy would certainly constitute a very serious experimental challenge, but a result would be highly significant.

\section*{Acknowledgments}
It is a pleasure to thank Gustavo Branco, Howie Haber, Gui Rebelo and Joao Silva for discussions.
BG is partially supported by the National Science Centre (Poland) under research project, decision no DEC-2011/01/B/ST2/00438. PO is supported by the Research Council of Norway.
%%%%%%%%%%%%%%%%%%%%%%%%%%%%%%%%%%%%%%%%%%%%%%%%%%%%%%%%%%%%%%%%%%%%%%%%%%%%%
\appendix
\section{Further properties of the model}
\label{Sec:MoreProperties}
\setcounter{equation}{0}
%%%%%%%%%%%%%%%%%%%%%%%%%%%%%%%%%%%%%%%%%%%%%%%%%%%%%%%%%%%%%%%%%%%%%%%%%%%%%
%
In this appendix we collect formulas relevant for minimization of the potential and
determination of scalar masses. We also show relations between quartic coupling constants
and masses and mixing angles, they are crucial to show that potential parameters could indeed be
expressed through the parameter set (\ref{input}) that we are adopting. It should be emphasized that
the relations presented here are applicable for the most general 2HDM without any $\zBB_2$
symmetry imposed on the model.
   
%%%%%%%%%%%%%%%%%%%%%%%%%%%%%%%%%%%%%%%%%%%%%%%%%%%%%%%%%%%%%%%%%%%%%%%%%%%%%
\subsection{Stationary points of the potential}
%%%%%%%%%%%%%%%%%%%%%%%%%%%%%%%%%%%%%%%%%%%%%%%%%%%%%%%%%%%%%%%%%%%%%%%%%%%%%
By demanding that the derivatives of the potential (\ref{Eq:pot}) with respect to the fields should vanish in the vacuum, we end up with the following stationary-point equations:
\begin{eqnarray}
m_{11}^2&=&v_1^2\lambda_1+v_2^2(\lambda_3+\lambda_4)+\frac{v_2^2}{c_\xi}(\Re\lambda_5 c_\xi-\Im\lambda_5s_\xi)\label{stationary1}\nonumber\\
&&+\frac{v_1v_2}{c_\xi}\left[\Re\lambda_6(2+c_{2\xi})-\Im\lambda_6s_{2\xi}\right]+\frac{v_2}{v_1c_\xi}(v_2^2\Re\lambda_7-\Re m_{12}^2),\\
m_{22}^2&=&v_2^2\lambda_2+v_1^2(\lambda_3+\lambda_4)+\frac{v_1^2}{c_\xi}(\Re\lambda_5c_\xi-\Im\lambda_5s_\xi)\label{stationary2}\nonumber\\
&&+\frac{v_1v_2}{c_\xi}\left[\Re\lambda_7(2+c_{2\xi})-\Im\lambda_7s_{2\xi}\right]+\frac{v_1}{v_2c_\xi}(v_1^2\Re\lambda_6-\Re m_{12}^2),\\
\Im m_{12}^2&=&\frac{v_1v_2}{c_\xi}(\Re\lambda_5s_{2\xi}+\Im\lambda_5c_{2\xi})+\frac{v_1^2}{c_\xi}(\Re\lambda_6 s_\xi+\Im\lambda_6 c_\xi)\nonumber\\
&&+\frac{v_2^2}{c_\xi}(\Re\lambda_7s_\xi+\Im\lambda_7c_\xi)-\Re m_{12}^2t_\xi,
\label{stationary3}
\end{eqnarray}
with $c_x=\cos x$, $s_x=\sin x$, and $t_x=\tan x$.
Thus, we may eliminate $m_{11}^2$, $m_{22}^2$ and $\Im m_{12}^2$ from the potential by these substitutions, thereby reducing the number of parameters of the model.
%%%%%%%%%%%%%%%%%%%%%%%%%%%%%%%%%%%%%%%%%%%%%%%%%%%%%%%%%%%%%%%%%%%%%%%%%%%%%
\subsection{The scalar masses and the re-expression of the $\lambda$s}
%%%%%%%%%%%%%%%%%%%%%%%%%%%%%%%%%%%%%%%%%%%%%%%%%%%%%%%%%%%%%%%%%%%%%%%%%%%%%
From the bilinear terms of potential, we may read off directly the mass of the charged scalars
\beq
M_{H^\pm}^2=\frac{v^2}{2v_1v_2c_\xi}\Re\left(m_{12}^2-v_1^2\lambda_6-v_2^2\lambda_7-v_1v_2e^{i\xi}\left[\lambda_4+\lambda_5\right]\right)\label{chargedmass},
\eeq
and the elements of the neutral-sector mass matrix
\begin{eqnarray}
{\cal M}_{11}^2&=&v_1^2\lambda_1-v_2^2s^2_\xi\Re\lambda_5-\frac{v_2^2}{2c_\xi}s_\xi c_{2\xi}\Im\lambda_5+\frac{v_1v_2}{2c_\xi}(1+2c_{2\xi})\Re\lambda_6\nonumber\\
&&-2v_1v_2s_\xi\Im\lambda_6-\frac{v_2^3}{2v_1c_\xi}\Re\lambda_7+\frac{v_2}{2v_1c_\xi}\Re m_{12}^2, \label{M11} \\
{\cal M}_{22}^2&=&v_2^2\lambda_2-v_1^2s^2_\xi\Re\lambda_5-\frac{v_1^2}{2c_\xi}s_\xi c_{2\xi}\Im\lambda_5+\frac{v_1v_2}{2c_\xi}(1+2c_{2\xi})\Re\lambda_7\nonumber\\
&&-2v_1v_2s_\xi\Im\lambda_7-\frac{v_1^3}{2v_2c_\xi}\Re\lambda_6+\frac{v_1}{2v_2c_\xi}\Re m_{12}^2, \label{M22} \\
{\cal M}_{33}^2&=&-v^2c^2_\xi\Re\lambda_5-\frac{v^2}{2c_\xi}(2s^3_\xi-3s_\xi)\Im\lambda_5\nonumber\\
&&-\frac{v^2v_1}{2v_2c_\xi}\Re\lambda_6-\frac{v^2v_2}{2v_1c_\xi}\Re\lambda_7+\frac{v^2}{2v_1v_2c_\xi}\Re m_{12}^2, \label{M33} \\
{\cal M}_{12}^2&=&v_1v_2(\lambda_3+\lambda_4)+v_1v_2c^2_\xi\Re\lambda_5+\frac{v_1v_2}{2c_\xi}(2s^3_\xi-3s_\xi)\Im\lambda_5+\frac{v_1^2}{2c_\xi}(2+c_{2\xi})\Re\lambda_6\nonumber\\
&&-v_1^2s_\xi\Im\lambda_6+\frac{v_2^2}{2c_\xi}(2+c_{2\xi})\Re\lambda_7-v_2^2s_\xi\Im\lambda_7-\frac{1}{2c_\xi}\Re m_{12}^2,  \label{M12} \\
{\cal M}_{13}^2&=&-\half vv_2s_{2\xi}\Re\lambda_5-\half vv_2c_{2\xi}\Im\lambda_5
-vv_1s_\xi\Re\lambda_6-vv_1c_\xi\Im\lambda_6,   \label{M13} \\
{\cal M}_{23}^2&=&-\half vv_1s_{2\xi}\Re\lambda_5-\half vv_1c_{2\xi}\Im\lambda_5
-vv_2s_\xi\Re\lambda_7-vv_2c_\xi\Im\lambda_7. \label{M23}
\end{eqnarray}
The eigenvalues of this matrix will be the masses of the three neutral scalars. In order to find these, a cubic equation needs to be solved. For our purposes, a different approach will suffice.
We may rewrite the elements of the mass matrix ${\cal M}_{ij}^2$ in terms of the eigenvalues $M_i^2$ and elements of the rotation matrix $R_{ij}$ as six equations:
\begin{eqnarray}
{\cal M}_{11}^2&=&M_1^2R_{11}^2+M_2^2R_{21}^2+M_3^2R_{31}^2, \label{M11eq} \\
{\cal M}_{22}^2&=&M_1^2R_{12}^2+M_2^2R_{22}^2+M_3^2R_{32}^2, \label{M22eq} \\
{\cal M}_{33}^2&=&M_1^2R_{13}^2+M_2^2R_{23}^2+M_3^2R_{33}^2,
\label{M33eq} \\
{\cal M}_{12}^2&=&M_1^2R_{11}R_{12}+M_2^2R_{21}R_{22}+M_3^2R_{31}R_{32}, \\
{\cal M}_{13}^2&=&M_1^2R_{11}R_{13}+M_2^2R_{21}R_{23}+M_3^2R_{31}R_{33},
\label{M13eq}\\
{\cal M}_{23}^2&=&M_1^2R_{12}R_{13}+M_2^2R_{22}R_{23}+M_3^2R_{32}R_{33}.
\label{M23eq}
\end{eqnarray}
We may now regard (\ref{chargedmass}) and (\ref{M11eq})--(\ref{M23eq}) as a set of seven equations. These equations are linear in the $\lambda_i$-parameters of the potential. We have 10 such parameters (counting both real and imaginary parts of $\lambda_5$, $\lambda_6$ and $\lambda_7$) and may now solve this set of seven equations for seven of the $\lambda_i$-parameters, thus expressing them in terms of the other parameters we have introduced. It is convenient to solve for the following set of parameters: $(\lambda_1,\lambda_2,\lambda_3,\lambda_4, {\rm Re}\lambda_5, {\rm Im}\lambda_6, {\rm Im}\lambda_7)$. We also introduce the more convenient parameter $\mu^2$ by putting
\bea
{\rm Re}\hspace*{0.1cm} m_{12}^2=\frac{2v_1v_2}{v^2}\mu^2.
\eea
Solving the set of equations, we arrive at
\bea
\lambda_1&=&
-\frac{v_2^2}{v^2 v_1^2c_\xi^3}\mu^2
+\frac{\left(R_{11} v-R_{13} v_2 t_\xi\right)^2}{v^2 v_1^2}M_1^2
+\frac{\left(R_{21} v-R_{23} v_2 t_\xi\right)^2}{v^2 v_1^2}M_2^2\nonumber\\
&&+\frac{\left(R_{31} v-R_{33} v_2t_\xi\right)^2}{v^2 v_1^2}M_3^2
-\frac{v_2^2 t_\xi}{2 v_1^2c_\xi^2}{\rm Im}\lambda_5\nonumber\\
&&-\frac{v_2 (2 c_{2\xi}+1)}{2 v_1c_\xi^3}{\rm Re}\lambda_6
+\frac{v_2^3}{2 v_1^3c_\xi^3}{\rm Re}\lambda_7,\label{l1rephrased}\\
\lambda_2&=&
-\frac{v_1^2}{v^2 v_2^2c_\xi^3}\mu^2
+\frac{\left(R_{12} v-R_{13} v_1 t_\xi\right)^2}{v^2 v_2^2}M_1^2
+\frac{\left(R_{22} v-R_{23} v_1 t_\xi\right)^2}{v^2 v_2^2}M_2^2\nonumber\\
&&+\frac{\left(R_{32} v-R_{33} v_1 t_\xi\right)^2}{v^2 v_2^2}M_3^2
-\frac{v_1^2 t_\xi}{2 v_2^2c_\xi^2}{\rm Im}\lambda_5\nonumber\\
&&+\frac{v_1^3}{2 v_2^3c_\xi^3}{\rm Re}\lambda_6
-\frac{v_1 (2c_{2\xi}+1)}{2 v_2c_\xi^3}{\rm Re}\lambda_7,\\
\lambda_3&=&
\frac{2}{v^2}M_{H^\pm}^2
-\frac{1}{v^2c_\xi^3}\mu^2
+\frac{\left(R_{12} v-R_{13} v_1 t_\xi\right) \left(R_{11} v-R_{13} v_2 t_\xi\right)}{v^2 v_1 v_2}M_1^2\nonumber\\
&&+\frac{\left(R_{22} v-R_{23} v_1 t_\xi\right) \left(R_{21} v-R_{23} v_2t_\xi\right)}{v^2 v_1 v_2}M_2^2\nonumber\\
&&+\frac{\left(R_{32} v-R_{33} v_1 t_\xi\right) \left(R_{31} v-R_{33} v_2t_\xi\right)}{v^2 v_1 v_2}M_3^2\nonumber\\
&&-\frac{1}{2 c_\xi^2} t_\xi {\rm Im}\lambda_5
-\frac{v_1 c_{2\xi} }{2 v_2c_\xi^3}{\rm Re}\lambda_6
-\frac{v_2 c_{2\xi} }{2 v_1c_\xi^3}{\rm Re}\lambda_7,\\
\lambda_4&=&
-\frac{2}{v^2}M_{H^\pm}^2
+\frac{c_{2\xi} }{v^2c_\xi^3}\mu^2
+\frac{R_{13}^2}{v^2c_\xi^2}M_1^2
+\frac{R_{23}^2}{v^2c_\xi^2}M_2^2
+\frac{R_{33}^2}{v^2c_\xi^2}M_3^2\nonumber\\
&&-\frac{1}{2c_\xi^2} t_\xi {\rm Im}\lambda_5
-\frac{v_1 c_{2\xi} }{2 v_2c_\xi^3}{\rm Re}\lambda_6
-\frac{v_2 c_{2\xi} }{2 v_1c_\xi^3}{\rm Re}\lambda_7,\\
{\rm Re}\lambda_5&=&
\frac{1}{v^2c_\xi^3}\mu^2
-\frac{R_{13}^2 }{v^2c_\xi^2}M_1^2
-\frac{R_{23}^2 }{v^2c_\xi^2}M_2^2
-\frac{R_{33}^2 }{v^2c_\xi^2}M_3^2\nonumber\\
&&+\frac{1}{4c_\xi^3} (3 s_\xi+s_{3\xi}) {\rm Im}\lambda_5
-\frac{v_1 }{2 v_2c_\xi^3}{\rm Re}\lambda_6
-\frac{v_2 }{2 v_1c_\xi^3}{\rm Re}\lambda_7,\\
{\rm Im}\lambda_6&=&
-\frac{v_2 t_\xi}{v^2 v_1c_\xi^2}\mu^2
+\frac{R_{13} \left(R_{13} v_2t_\xi-R_{11} v\right)}{v^2 v_1c_\xi}M_1^2
+\frac{R_{23} \left(R_{23} v_2 t_\xi-R_{21} v\right)}{v^2 v_1c_\xi}M_2^2\nonumber\\
&&+\frac{R_{33} \left(R_{33} v_2 t_\xi-R_{31} v\right)}{v^2 v_1c_\xi}M_3^2
-\frac{v_2 }{2 v_1c_\xi^3}{\rm Im}\lambda_5
-\frac{1}{2c_\xi^2}t_\xi c_{2\xi}{\rm Re}\lambda_6
+\frac{v_2^2 t_\xi}{2 v_1^2c_\xi^2}{\rm Re}\lambda_7,\nonumber\\ \\
{\rm Im}\lambda_7&=&
-\frac{v_1 t_\xi}{v^2 v_2c_\xi^2}\mu^2
+\frac{R_{13}  \left(R_{13} v_1 t_\xi-R_{12} v\right)}{v^2 v_2c_\xi}M_1^2
+\frac{R_{23}  \left(R_{23} v_1 t_\xi-R_{22} v\right)}{v^2 v_2c_\xi}M_2^2\nonumber\\
&&+\frac{R_{33}  \left(R_{33} v_1 t_\xi-R_{32} v\right)}{v^2 v_2c_\xi}M_3^2
-\frac{v_1 }{2 v_2c_\xi^3}{\rm Im}\lambda_5
+\frac{v_1^2 t_\xi}{2 v_2^2c_\xi^2}{\rm Re}\lambda_6
-\frac{1}{2c_\xi^2}t_\xi c_{2\xi}{\rm Re}\lambda_7.\label{iml7rephrased}\nonumber\\
\eea
These substitutions enable us to express quantities (couplings, observables, etc.) arising from the vevs and the potential in terms of the set of parameters $\pcal_{67}$ of equation~(\ref{input}).
We also note that each of the seven expressions listed above is linear in the subset of parameters denoted $\pcal_0$ and given by equation~(\ref{eq:pcal0}).

%%%%%%%%%%%%%%%%%%%%%%%%%%%%%%%%%%%%%%%%%%%%%%%%%%%%%%%%%%%%%%%%%%%%%%%%%%%%%
\section{Relevant coupling coefficients}
\label{Sec:Couplings}
\setcounter{equation}{0}
%%%%%%%%%%%%%%%%%%%%%%%%%%%%%%%%%%%%%%%%%%%%%%%%%%%%%%%%%%%%%%%%%%%%%%%%%%%%%
The couplings involving scalars can be read off from the relevant parts of the Lagrangian, and hence the Feynman rules for 
different interactions can be found. We shall here present expressions for the 
couplings relevant for the bosonic sector of the model. Some of them were not adopted explicitly in the 
main text, however we find it useful to collect them here for completeness and future reference.
The couplings involving physical scalars only are quite lengthy, so we will 
introduce the following abbreviations in order to compactify them:
\bea
e_i &\equiv& v_1R_{i1}+v_2R_{i2},\label{eq:e_i-def}\\
f_i&\equiv& v_1R_{i2}-v_2R_{i1}-ivR_{i3},\label{eq:f_i-def}\\
g_i&\equiv& v_1^3R_{i2}+v_2^3R_{i1},\\
h_{iik}&\equiv& v_1R_{i2}^2R_{k2}+v_2R_{i1}^2R_{k1}.
\eea
%%%%%%%%%%%%%%%%%%%%%%%%%%%%%%%%%%%%%%%%%%%%%%%%%%%%%%%%%%%%%%%%%%%%%%%%%%%%%
\subsection{The $H_iH^-H^+$, $H_iH_iH^-H^+$ and $H^-H^-H^+H^+$ couplings}
%%%%%%%%%%%%%%%%%%%%%%%%%%%%%%%%%%%%%%%%%%%%%%%%%%%%%%%%%%%%%%%%%%%%%%%%%%%%%
In addition to containing terms bilinear in the fields (that give us the masses of the scalar particles), the potential also contains trilinear and quadrilinear terms corresponding to interactions between the fields. 
In the present work, we shall need the $H_iH^-H^+$, $H_iH_iH^-H^+$ and $H^-H^-H^+H^+$ couplings. Reading the coefficients of these interactions directly off from the potential, we find
\bea
q_{i}&\equiv&\text{Coefficient}(V,H_iH^-H^+)\label{eq:qi}
\nonumber\\
&=&\frac{2 e_i}{v^2}M_{H^\pm}^2
-\frac{R_{i2} v_1+R_{i1} v_2-R_{i3} v t_\xi}{v_1 v_2c_\xi}\mu^2
+\frac{g_i-R_{i3} v^3 t_\xi}{v^2 v_1 v_2}M_i^2
+\frac{R_{i3} v^3}{2 v_1 v_2c_\xi^2}\Im\lambda_5\nonumber\\
&&-\frac{v^2 \left(R_{i3} v t_\xi-R_{i2} v_1+R_{i1} v_2\right)}{2 v_2^2c_\xi}\Re\lambda_6
-\frac{v^2 \left(R_{i3} v t_\xi+R_{i2} v_1-R_{i1} v_2\right)}{2 v_1^2c_\xi}\Re\lambda_7,\\
%%%%
q_{ii}&\equiv&\text{Coefficient}(V,H_iH_iH^-H^+)\label{eq:qii}
\nonumber\\
&=&\frac{e_i^2}{v^4}M_{H^\pm}^2
-\frac{ 
\left(v_1^2-v_2^2\right)^2|f_i|^2 + 2 v_1 v_2 e_ig_i
-2 v^3 v_1 v_2 e_i R_{i3}t_\xi
+v^4|f_i|^2t_\xi^2
}{2 v^4 v_1^2 v_2^2c_\xi }\mu^2\nonumber\\
&&+\frac{1}{2 v^4 v_1^2 v_2^2}\sum_{k=1}^3
\left[
g_k \left(2 e_i R_{i3} R_{k3} v_1 v_2+g_k R_{i3}^2+h_{iik} v^2\right)\right.\nonumber\\
&&\hspace*{3cm}
-vR_{k3}\left(g_k \left(|f_i|^2+R_{i3}^2 v^2\right)+2 e_i R_{i3} R_{k3} v_1 v_2 v^2+h_{iik} v^4\right) t_\xi\nonumber\\
&&\hspace*{3cm}\left.
+|f_i|^2 R_{k3}^2 v^4 t_\xi^2
\right] M_k^2
\nonumber\\
&&+\frac{2 e_i R_{i3} v v_1 v_2-v^2|f_i|^2 t_\xi}{4 v_1^2 v_2^2c_\xi^2}
\Im\lambda_5\nonumber\\
&&+\frac{1}{4 v_1 v_2^3c_\xi}
\left[
v^2 \left(R_{i3}^2 \left(v_1^2-v_2^2\right)+R_{i2}^2 v_1^2-R_{i1}^2 v_2^2\right)-2 |f_i|^2 v_2^2\right.\nonumber\\
&&\hspace*{2cm}\left.
-2 e_i R_{i3} v v_1 v_2 t_\xi
+|f_i|^2 v^2  t_\xi^2
\right]\Re\lambda_6\nonumber\\
&&+\frac{1}{4 v_2 v_1^3c_\xi}
\left[
-v^2 \left(R_{i3}^2 \left(v_1^2-v_2^2\right)+R_{i2}^2 v_1^2-R_{i1}^2 v_2^2\right)-2|f_i|^2 v_1^2\right.\nonumber\\
&&\hspace*{2cm}\left.
-2 e_i R_{i3} v v_1 v_2 t_\xi
+v^2|f_i|^2 t_\xi^2
\right]\Re\lambda_7,\\
q&\equiv&\text{Coefficient}(V,H^-H^-H^+H^+)\label{eq:q}
\nonumber\\
&=&
-\frac{1}{2 v^2 v_1^2 v_2^2c_\xi}
\left[
\left(v_1^2-v_2^2\right)^2+v^4 t_\xi^2
\right]\mu^2
+\sum_{k=1}^3\frac{\left(g_k-R_{k3} v^3 t_\xi\right)^2}{2 v^4 v_1^2 v_2^2}M_k^2
-\frac{v^4 t_\xi }{4 v_1^2 v_2^2c_\xi^2}\Im\lambda_5\nonumber\\
&&
+\frac{v^2\left(
v_1^2-3 v_2^2+v^2t_\xi^2
\right)}{4 v_1 v_2^3c_\xi}
\Re\lambda_6
+\frac{v^2\left(v_2^2-3 v_1^2+v^2 t_\xi^2\right)}{4 v_2 v_1^3c_\xi}\Re\lambda_7.
\label{Eq:coupling-q}
\eea
An important property of these couplings is that they are linear in the parameters of the set $\pcal_0$.
It is also worth pointing out that these couplings are not identical to the Feynman rules for the interactions they represent. One needs to take into account the fact that the potential appears with a negative sign in the Lagrangian as well as the fact that combinatorial factors arising from the presence of multiple identical particles, and the imaginary unit $i$ should be present in the interactions terms. The corresponding Feynman rules become
\bea
H_iH^+H^-:&& -iq_i,\\
H_iH_iH^+H^- && -2iq_{ii},\\
H^+H^+H^-H^- && -4iq.
\eea
There are many interesting and useful relations among the couplings of the model. We will point out a couple of these, but first let us introduce a quantity that is completely symmetric under the interchange of any two of the indices $1,2,3$:
\bea
\sigma\equiv q_{11}+q_{22}+q_{22}\label{eq:sigma}.
\eea
One can now easily show that 
\bea
e_iq_i=(\sigma-q)e_i^2+(q_{ii}-q)v^2,
\eea
which in turn leads to 
\bea
e_1q_1+e_2q_2+e_3q_3=2(\sigma-2q)v^2,
\eea
where we have used the fact that $e_1^2+e_2^2+e_3^2=v^2$.
%%%%%%%%%%%%%%%%%%%%%%%%%%%%%%%%%%%%%%%%%%%%%%%%%%%%%%%%%%%%%%%%%%%%%%%%%%%%%
\subsection{Couplings containing the factor $e_i$}
%%%%%%%%%%%%%%%%%%%%%%%%%%%%%%%%%%%%%%%%%%%%%%%%%%%%%%%%%%%%%%%%%%%%%%%%%%%%%
The quantity $e_i = v_1R_{i1}+v_2R_{i2}$ plays an important role as it appears as a factor in numerous interactions involving neutral scalars.
Here, we list them in two groups, those that contain the antisymmetric $\epsilon_{ijk}$, and those that do not:
\begin{equation} \label{eq:e_i-epsilon}
H_i H_j Z_\mu: \quad
\frac{g}{2v\cos\thetaW}\epsilon_{ijk}e_k(p_i-p_j)_\mu, \quad
H_i H_j G_0:\quad 
i\frac{M_i^2-M_j^2}{v^2}\epsilon_{ijk}e_k,
\end{equation}
and
\begin{subequations}
\begin{alignat}{2}
H_i Z_\mu Z_\nu:\quad &
\frac{ig^2}{2\cos^2\thetaW}e_i\,g_{\mu\nu}, &\quad
H_i W^+_\mu W^-_\nu:\quad &
\frac{ig^2}{2}e_i\,g_{\mu\nu}, \label{eq:H_iZZ}\\
H_i G_0 G_0:\quad &
\frac{-iM_i^2 e_i}{v^2}, &\quad
H_i G^+ G^-:\quad &
\frac{-iM_i^2 e_i}{v^2},\\
H_i G^+ A_\mu W^-_\nu: \quad &
\frac{ig^2\sin\thetaW}{2v}e_i \,g_{\mu\nu}, &\quad
H_i G^- A_\mu W^+_\nu: \quad &
\frac{ig^2\sin\thetaW}{2v}e_i \,g_{\mu\nu}, \\
H_i G^+ Z_\mu W^-_\nu: \quad &
-\frac{ig^2}{2v} \frac{\sin^2\thetaW}{\cos\thetaW}e_i \,g_{\mu\nu}, &\quad
H_i G^- Z_\mu W^+_\nu: \quad &
-\frac{ig^2}{2v} \frac{\sin^2\theta_\text{W}}{\cos\thetaW}e_i \,g_{\mu\nu}, \\
H_i G_0 Z_\mu: \quad &
\frac{g}{2v\cos\thetaW}e_i(p_i-p_0)_\mu, \\
H_i G^+ W^-_\mu: \quad &
i\frac{g}{2v}e_i(p_i-p^+)_\mu, &\quad
H_i G^- W^+_\mu: \quad &
-i\frac{g}{2v}e_i(p_i-p^-)_\mu.
\end{alignat}
\end{subequations}
Here, $G_0$ and $G^\pm$ denote the Goldstone fields.
Since the couplings (\ref{eq:e_i-epsilon}) are the {\it only} ones that contain the antisymmetric $\epsilon_{ijk}$, one of these vertices must be involved in each of the invariants $\Im J_1$, $\Im J_2$,  and $\Im J_{30}$.

In the notation of \cite{Accomando:2006ga,El_Kaffas:2006nt}, the different values can be expressed as
\begin{subequations}
\begin{align}
e_1&=v\cos\alpha_2\cos(\beta-\alpha_1), \\
e_2&=v[\cos\alpha_3\sin(\beta-\alpha_1)-\sin\alpha_2\sin\alpha_3\cos(\beta-\alpha_1)], \\
e_3&=-v[\sin\alpha_3\sin(\beta-\alpha_1)+\sin\alpha_2\cos\alpha_3\cos(\beta-\alpha_1)].
\end{align}
\end{subequations}
In the limits of CP conservation that are not a consequence of mass degeneracy \cite{ElKaffas:2007rq}, one of the $e_i$ will vanish. For the 2HDM5, they are given as:
\begin{subequations}
\begin{alignat}{4}
H_1&=A:&\quad e_1&=0,&\quad e_2&=v\sin(\beta-\alpha_1\mp\alpha_3), 
&\quad e_3&=\mp v\cos(\beta-\alpha_1\mp\alpha_3), \\
H_2&=A:&\quad e_1&=v\cos(\beta-\alpha_1), &\quad e_2&=0, 
&\quad e_3&=-v\sin(\beta-\alpha_1),\\
H_3&=A:&\quad e_1&=v\cos(\beta-\alpha_1), &\quad e_2&=v\sin(\beta-\alpha_1), 
&\quad e_3&=0,
\end{alignat}
\end{subequations}
with $A$ here denoting the CP-odd neutral Higgs boson (not the photon).
Substituting the mixing angle $\alpha$ describing the CP-conserving case, $\alpha_1=\alpha+\pi/2$, we recover the familiar $H_1ZZ$ and $H_1W^+W^-$ couplings (see Eq.~(\ref{eq:H_iZZ})) proportional to $e_1=v\sin(\beta-\alpha)$, valid when $H_2$ or $H_3$ is odd under $CP$.

We note that
\begin{equation}
e_1^2+e_2^2+e_3^2=v^2.
\end{equation}
Clearly, equipartition maximizes the product $e_1e_2e_3$ which enters in $\Im J_2$. Thus
\begin{equation}
\max\frac{e_1e_2e_3}{v^3}=\frac{1}{3\sqrt{3}}\simeq0.1925.
\end{equation}

%%%%%%%%%%%%%%%%%%%%%%%%%%%%%%%%%%%%%%%%%%%%%%%%%%%%%%%%%%%%%%%%%%%%%%%%%%%%%
\subsection{The coupling $q_i$}
%%%%%%%%%%%%%%%%%%%%%%%%%%%%%%%%%%%%%%%%%%%%%%%%%%%%%%%%%%%%%%%%%%%%%%%%%%%%%
The trilinear neutral-charged Higgs coupling is denoted as $q_i$:
\begin{equation}
H_i H^+H^-:\quad 
-iq_i.
\end{equation}
This coupling is more complicated than $e_i$ and $f_i$, when expressed in terms of the neutral-sector mixing matrix.

In the model discussed in \cite{Grzadkowski:2013rza}, the coupling coefficient $q_i$ takes the form 
\begin{align}
H_iH^+H^-: \quad
& -i q_i  \\
=&-i\biggl[2M_{H^\pm}^2\frac{v_1R_{i1}+v_2R_{i2}}{v^2}
-\mu^2\frac{v_1R_{i2}+v_2R_{i1}}{v_1v_2}
+M_i^2\frac{v_1^3R_{i2}+v_2^3R_{i1}}{v_1v_2v^2}
+\Delta\frac{v R_{i3}}{v_1v_2}\biggr]. \nonumber
\end{align}
Here, $\Delta$ can be written as $\Delta\equiv v\Delta_{123}$, with
\begin{equation}
\Delta_{ijk}=\frac{(M_k^2-M_j^2)R_{j3}R_{k3}}{v_1R_{i1}-v_2R_{i2}}.
\end{equation}

%%%%%%%%%%%%%%%%%%%%%%%%%%%%%%%%%%%%%%%%%%%%%%%%%%%%%%%%%%%%%%%%%%%%%%%%%%%%%
\subsection{The coupling $f_i$}
%%%%%%%%%%%%%%%%%%%%%%%%%%%%%%%%%%%%%%%%%%%%%%%%%%%%%%%%%%%%%%%%%%%%%%%%%%%%%
Another coefficient appearing in many bosonic couplings is denoted $f_i$,
\begin{equation}
f_i\equiv v_1R_{i2}-v_2R_{i1}-ivR_{i3}.
\end{equation}
In contrast to the $e_i$ (and $q_i$) it is complex. 
However, we note that they are related to the $e_i$ as follows:
\begin{equation}
\Re f_i f_j^\ast=v^2\delta_{ij}-e_i e_j.
\end{equation}

\begin{subequations}
\begin{alignat}{2}
H_i G^-H^+:\quad &
-i\frac{M_i^2-M_{H^\pm}^2}{v^2}f_i, &\quad
H_i G^+H^-:\quad &
-i\frac{M_i^2-M_{H^\pm}^2}{v^2}f_i^\ast, \label{eq:H_iGH}\\
H_iH^+A_\mu W^{-\mu}: \quad &
i\frac{g^2}{2v}\sin\thetaW f_i, &\quad
H_iH^-A_\mu W^{+\mu}: \quad &
i\frac{g^2}{2v}\sin\thetaW f_i^\ast, \\
H_i H^+ Z_\mu W^{-\mu}: \quad &
\frac{-ig^2}{2v}\frac{\sin^2\thetaW}{\cos\thetaW} f_i, &\quad
H_i H^- Z_\mu W^{+\mu}: \quad &
\frac{-ig^2}{2v}\frac{\sin^2\thetaW}{\cos\thetaW} f_i^\ast, \\
H_i H^- W^{+\mu}: \quad &
\frac{ig}{2v} f_i (p_i-p^-)^\mu, &\quad
H_i H^+ W^{-\mu}: \quad &
\frac{-ig}{2v} f_i^\ast (p_i-p^+)^\mu.
\end{alignat}
\end{subequations}
In the CP-conserving limit, two of the $f_i$ are real, whereas the third is pure imaginary.

We note that the following relation follows from the unitarity of the rotation matrix:
\begin{equation}
\sum_{i,j,k}\epsilon_{ijk} e_i f_j^\ast f_k = -2iv^3.
\end{equation}

%%%%%%%%%%%%%%%%%%%%%%%%%%%%%%%%%%%%%%%%%%%%%%%%%%%%%%%%%%%%%%%%%%%%%%%%%%%%%%%

\end{document}